\documentclass[12pt,a4paper]{scrartcl}
\def\version{arxiv}

\usepackage{a4wide}
\usepackage{graphicx}
\usepackage{url}
\usepackage[round,colon]{natbib}
\usepackage{listings}
\usepackage{caption}
\usepackage{subcaption}
\usepackage{units}
\usepackage{booktabs}
\usepackage{textcomp}
\usepackage[backgroundcolor=white,textsize=small]{todonotes}
\usepackage{amsmath}
\usepackage{xspace}
\usepackage{array}
\usepackage{times}
\usepackage{ifthen}

\usepackage[modulo]{lineno}

\usepackage{sectsty}
\allsectionsfont{ \large }

\usepackage{verbatim}
\newcommand\wordcount{\verbatiminput{main.sum}}

\usepackage{hyperref}
\hypersetup{
  colorlinks=true,
  urlcolor=blue,
  linkcolor=blue,
  citecolor=blue
}

\usepackage{manyfoot}
\DeclareNewFootnote{NF}

\DeclareNewFootnote{SF}

\newcommand{\synapse}{{\sc synapse}\xspace}

\newcommand{\waferscale}{BrainScaleS wafer-scale system\xspace}
\newcommand{\emp}{EPP\xspace}
\ifthenelse{\equal{\version}{frontiers}}{
  \newcommand{\figw}{85mm}
  \newcommand{\figww}{180mm}
}{
  \newcommand{\figw}{85mm}
  \newcommand{\figww}{\textwidth}
}
\newcommand{\mansubref}[1]{(#1)\xspace} 
\newcommand{\manref}[2]{\ref{#1}#2\xspace}

\makeatletter
\renewcommand*{\@seccntformat}[1]{%
   \csname the#1\endcsname.\quad
}
\makeatother

\newcommand{\greatTitle}{Reward-based learning under hardware constraints - Using a RISC processor embedded in a neuromorphic substrate}
\title{\greatTitle}

\author{
Simon Friedmann$^{1*}$, \\ 
Nicolas Fremaux$^{2*}$, \\ 
Johannes Schemmel$^{1}$,\\
Wulfram Gerstner$^{2}$, \\
Karlheinz Meier$^{1}$ \\
\vspace{5pt}\\
$^{1}$Kirchhoff Institute for Physics, University of Heidelberg, Germany\\
$^{2}$School of Computer and Communication Sciences and Brain-Mind Institute, Ecoles Polytechniques Federales de Lausanne, Switzerland
}

\begin{document}

\onecolumn
\rm

\begin{center}
	{\bf \Large \greatTitle}\\
\vspace{20pt}
{\bf Simon Friedmann$^{\text{*}1}$, 
Nicolas Fr\'emaux$^{2}$, 
Johannes Schemmel$^{1}$,
Wulfram Gerstner$^{2}$,
Karlheinz Meier$^{1}$}
\end{center}
\begin{flushleft}
$^{1}$~Kirchhoff Institute for Physics, Ruprecht-Karls-University Heidelberg, Germany\\
$^{2}$~School of Computer and Communication Sciences and Brain-Mind Institute, Ecole Polytechnique F\'ed\'erale de Lausanne, Switzerland\\
\end{flushleft}
\vspace{15pt}

\ifthenelse{\equal{\version}{frontiers}}{
  {\bf \noindent Running title:}\\
  Reward-based learning under hardware constraints\\
  \vspace{10pt}

  {\bf \noindent Word count}\\
  \wordcount
}{}

\vfill

{\bf \noindent $^\text{*}$Correspondence:}\\
Simon Friedmann\\
Ruprecht-Karls-University Heidelberg\\
Kirchhoff Institute for Physics\\
Im Neuenheimer Feld 227\\
69120 Heidelberg, Germany\\
simon.friedmann@kip.uni-heidelberg.de\\
\vspace{20pt}

\newpage


{\bf \large \noindent Abstract}\\

\noindent In this study, we propose and analyze in simulations a new, highly flexible method of implementing synaptic plasticity in a wafer-scale, accelerated neuromorphic hardware system.
The study focuses on globally modulated STDP, as a special use-case of this method.
Flexibility is achieved by embedding a general-purpose processor dedicated to plasticity into the wafer.
To evaluate the suitability of the proposed system, we use a reward modulated STDP rule in a spike train learning task.
A single layer of neurons is trained to fire at specific points in time with only the reward as feedback.
This model is simulated to measure its performance, i.e.\ the increase in received reward after learning.
Using this performance as baseline, we then simulate the model with various constraints imposed by the proposed implementation and compare the performance.
The simulated constraints include discretized synaptic weights, a restricted interface between analog synapses and embedded processor, and mismatch of analog circuits.
We find that probabilistic updates can increase the performance of low-resolution weights, a simple interface between analog synapses and processor is sufficient for learning, and performance is insensitive to mismatch.
Further, we consider communication latency between wafer and the conventional control computer system that is simulating the environment.
This latency increases the delay, with which the reward is sent to the embedded processor.
Because of the time continuous operation of the analog synapses, delay can cause a deviation of the updates as compared to the not delayed situation.
We find that for highly accelerated systems latency has to be kept to a minimum.
This study demonstrates the suitability of the proposed implementation to emulate the selected reward modulated STDP learning rule.
It is therefore an ideal candidate for implementation in an upgraded version of the wafer-scale system developed within the BrainScaleS project.


\vfill


\newpage
\ifthenelse{\equal{\version}{frontiers}}{
  \linenumbers
}{}
\section{Introduction}
In reinforcement learning, an agent learns to achieve a goal through interaction with an environment~\citep{sutton1998reinforcement}.
The environment provides a single scalar number, the reward, as feedback to the actions performed by the learning agent.
The agent tries to maximize the reward it receives over time by changing its behavior.
In contrast to supervised learning, where an instructor supplies the correct actions to take, here the agent has to learn the correct strategy itself through trial-and-error.
Typically this is done by introducing randomness in the selection of actions and taking into account the resulting reward.
Recently, several studies have suggested extending classical spike-timing dependent plasticity \citep[STDP,][]{morrison08_stdp,caporale08_stdp} into reward-modulated STDP to implement reinforcement learning in the context of spiking neural networks \citep{Izhikevich2007,Farries2007,Florian2007,legenstein2008learning,Fremaux2010,Potjans2011}.
One of the key issues in reinforcement learning is solving the so-called temporal credit assignment problem:
reward arrives some time after the action that caused it.
So how does the agent know how to change its behavior?
It needs to retain some information about recent actions in order to assign proper credit for the rewards it receives.
To do this, reward modulated STDP generates an eligibility trace for every synapse that depends on pre- and postsynaptic firing.
This trace, modulated by the reward, determines the change of synaptic weight, thereby solving the credit assignment problem.

Spike-based implementations do not only offer an approach to biological models of learning, they are also suitable for implementation in neuromorphic hardware devices.
Existing systems offer a number of interesting characteristics, such as low-power consumption \citep[e.g.][]{wijekoon2008,livi09,Seo2011}, faster than real-time dynamics \citep{wijekoon2008,schemmel_iscas2010}, and scalability to large networks \citep{schemmel_iscas2010,furber2012}.
They are typically built with two goals in mind:
as new kind of brain inspired information processing device and to provide a scalable platform for the experimental exploration of networks.
Several studies so far have focused on the implementation of variants of unsupervised STDP in neuromorphic hardware \citep{indiveri_tnn2006,schemmel_ijcnn06,Seo2011,ramakrishnan2011floating,Davies20123}.
The synapse circuit presented by \cite{Wijekoon2011} implements the model proposed by \cite{Izhikevich2007} with the goal of enabling reward modulated STDP.

In this study we analyze the implementability of a reward modulated STDP model derived from \cite{Fremaux2010} as one example of a flexible hardware learning system.
To that end, we propose an extended version of the \waferscale \citep{schemmel_ijcnn2008,fieres_ijcnn2008,schemmel_iscas2010} to serve as a conceptual basis for this analysis.
This system is designed as a faster than real-time and flexible emulation platform for large neural networks.
The use of specialized analog circuits promises a higher power-efficiency than conventional digital simulations on supercomputers \citep{mead90neuromorphic}.
The acceleration in time compared to biology also makes the system interesting for reinforcement learning, which typically suffers from slow convergence \citep{sutton1998reinforcement}.
Starting from an existing system with limited modifications leads to a more realistic design prototype compared to starting from scratch.

A key objective for the proposed neuromorphic system is to be a valuable tool for neuroscience.
Therefore, the design must not be targeted at a single network architecture, task or learning rule, but instead stay as flexible as is reasonably possible.
On the other hand, implementing large-scale neural networks with accelerated time-scale raises technical challenges and trade-offs have to be made between flexibility and performance.
The proposed extension represents a plasticity mechanism reflecting this design philosophy:
specialized analog circuits in every synapse are combined with a general purpose embedded plasticity-processor (\emp).
This way, the benefits from the worlds of analog and processor-based computing can be combined:
analog circuits are used for compact, power-efficient and fast local processing, and digital processors allow for programmable plasticity rules.
Integrating the processors into the same application specific integrated circuits (ASIC) on the wafer as the neuromorphic substrate allows for scalability to wafer size networks and beyond.

In the following, we will consider only the aforementioned rule studied in \cite{Fremaux2010} and analyze effects caused by the adaptation to the hardware system in simulations.
We want to answer the question whether the hybrid approach combining processor and analog circuits is a suitable platform for this particular learning rule.
Among the hardware-induced constraints are non-continuous weights, drift of analog circuits and communication latency between hardware substrate and the controlling computer system.
We want to test and compare the performance of the unconstrained and the constrained plasticity rules in order to find guidelines for the hardware implementation, for example the required weight resolution or maximum noise levels.
Section~\ref{sec:mat_meth} describes the extended hardware system and the plasticity model.
Section~\ref{sec:results} presents results from simulations showing performance under hardware constraints.
Section~\ref{sec:discussion} provides a discussion of our results.

\section{Materials and methods}
\label{sec:mat_meth}
\subsection{Using an embedded processor for plasticity}
The key concept of our hardware implementation of synaptic plasticity is to use a programmable general-purpose processor in combination with fixed-function analog hardware.
Software running on the processor can use observables and controls to interface with the neuromorphic substrate.
Thereby, it is possible to flexibly switch between synaptic learning rules or use different ones in parallel for different synapses.
The alternative to this concept would be to use fixed-function hardware instead of a general-purpose processor.
This would allow a more efficient implementation of one specific rule, at the cost of system versatility.
In the following, we give background information on a complete neuromorphic system following the concept of processor-enabled plasticity.
From the system described, we derive hardware constraints that are used in the simulations reported in Section~\ref{sec:results}.

\subsubsection{System overview}
\begin{figure}
  \centering
  \includegraphics[width=\figw]{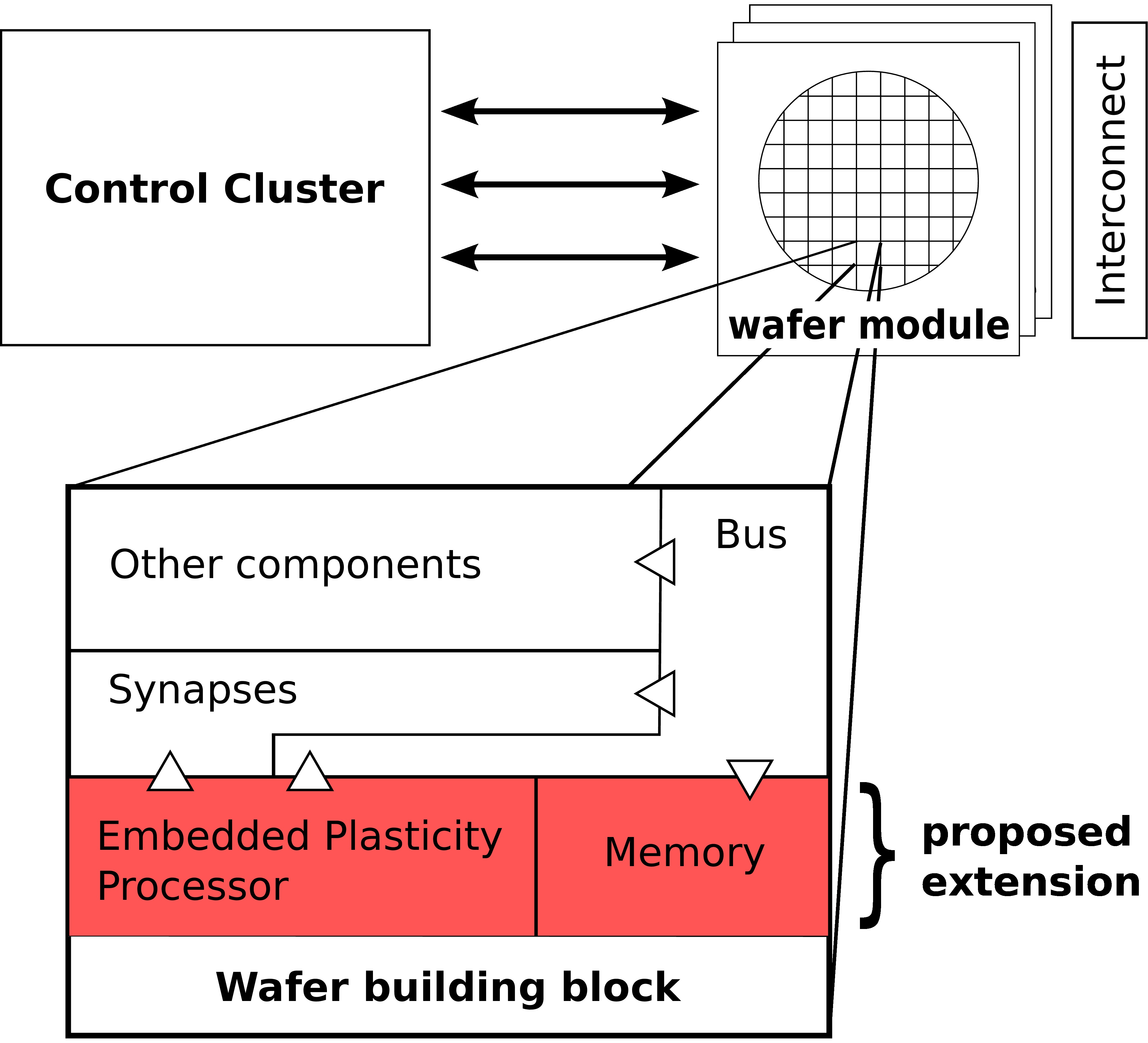}
  \caption[]{
    Overview of the system.
    The user controls the system through a cluster of conventional computers by sending configuration and spike data to a number of modules that each carry a wafer.
    These wafer modules are interconnected with a high-speed network to exchange spike events.
    The wafer contains identical building blocks, of which one is shown in an expanded view. 
    The proposed extension to the \waferscale in form of the embedded plasticity processor is marked in red.
    Input/output access from the processor to other components of the building block is indicated with triangles.
  }
  \label{fig:systemoverview}
\end{figure}

Figure~\ref{fig:systemoverview} gives a schematic overview of the complete hardware system.
The experimenter controls the system through a control cluster of off-the-shelf computers.
The network is provided in a description abstracted from the details of the system using the PyNN modelling language \citep{davison08pynn}.
An automated mapping process translates the description into the detailed configuration that is written to the wafer-modules \citep{ehrlich2010anniip,wendt08}.
These modules are interconnected by a high-speed network to communicate spike-events \citep{scholze11b}.
External stimulation can be applied to the network from the control cluster, using the high-speed links that are also used for configuration.
The wafer itself is subdivided into building blocks that contain the neuromorphic substrate, i.e.\ synapses, neurons, parameter storage and networking resources for spike transmission.

Our proposed extension adds an embedded plasticity processor (\emp) to every building block on the wafer, together with its own memory for instructions and data.
It will be equipped with three interfaces to the fixed-function hardware:
read and write access on the synapses, rate counters and event generation for the network and access to the control bus of the building block.
The latter is also used by external control accesses and thus, a plasticity program running on the embedded processor will be able to do everything that could be done from an off-wafer control computer as long as it only requires information local to the block.
There is no direct communication channel between processors envisioned, but software on the control computer could be used for data exchange.

\subsubsection{Implementing plasticity}
\label{sec:implementing}

Our proposed design represents a hybrid system, in which the digital \emp interacts closely with analog components.
Every synapse contains an analog accumulation circuit, similar to the version used in an earlier design \citep{schemmel_iscas07}.
For each pre-post and post-pre spike-pair, the time difference $\Delta t$ is measured and weighted exponentially using the amplitude $A_\pm$ and time constant $\tau_\pm$:
\begin{equation}
\delta_\pm = A_\pm \exp\left(-\frac{|\Delta t|}{\tau_\pm}\right).
\label{eqn:hwdeltae}
\end{equation}
These values are added to two local capacitors $a_+$ and $a_-$, respectively.
In the extended version the \emp will select synapses for readout and use an analog evaluation unit to produce a series of bits $b_i$ out of $a_+$ and $a_-$.
The evaluation function can perform different readout operations controlled by configuration bits $e_{cc}^i$, $e_{ca}^i$, $e_{ac}^i$ and $e_{aa}^i$ and analog parameters $a_{tl}$ and $a_{th}$:
\begin{equation}
  b_i = 
  \begin{cases}
    1 & \text{if } \frac{a_{tl}+e_{ac}^i a_+ + e_{ca}^i a_-}{1+e_{ac}^i +e_{ca}^i } > \frac{a_{th} + e_{cc}^i a_+ + e_{aa}^i  a_-}{1+e_{cc}^i +e_{aa}^i } \\
    0 & \text{otherwise}
  \end{cases}.
  \label{eqn:eval}
\end{equation}
Using $b_0\ldots b_{N-1}$, the current weight of the synapse $w$ and possibly further global parameters $P_0\ldots P_{M-1}$ as input, the weight update $\Delta$ is then calculated in software by the \emp:
\begin{equation}
  \Delta = F\left( b_0,\dots,b_{N-1},w,P_0,\dots,P_{M-1} \right)
  \label{eqn:update}
\end{equation}
Then, the new weight $w' = w + \Delta$ is written to weight storage by the plasticity program.
Using two evaluations $b_0,b_1$ with different sets of configuration bits, a simple example for $F$ would be:
\begin{equation}
  F\left( b_0, b_1 \right) = \tilde{A}_0 b_0 + \tilde{A}_1 b_1
  \label{eqn:updateex}
\end{equation}
With arbitrary constants $\tilde{A}_0$ and $\tilde{A}_1$.

Synapses in the system are organized in an array of synapse-units, where each synapse has a \unit[4]{bit} weight memory implemented with static random-access memory (SRAM) cells.
These offer the ability to combine adjacent units to increase resolution to \unit[8]{bit}.
Of course this has the negative effect of reducing the total amount of implementable synapses.

\subsubsection{Embedded micro-processor}
\label{sec:emp}
Plasticity algorithms will be implemented by software programs executed on the \emp.
A large class of micro-processors is in use today for various different applications from supercomputers, to smartphones and embedded controllers for traffic lights.
They all use different computer architectures reflecting the specific requirements and constraints of their application.

There are three important characteristics for a processor:
one, the used instruction set architecture (ISA) that defines coding and semantics of instructions and registers.
Two, whether instructions are executed out-of-order and three, whether the design is super-scalar, i.e.\ instructions can execute in parallel.
The instruction set architecture used here is a subset of the PowerISA 2.06 specification for \unit[32]{bit} \citep{powerisa_206}.
The main reason to use an existing ISA is the availability of compilers and tools.
Code for the \emp can be generated using the GNU Compiler Collection \citep{Stallman}, using the C programming language.

\begin{figure}
  \centering
  \includegraphics[width=\figw]{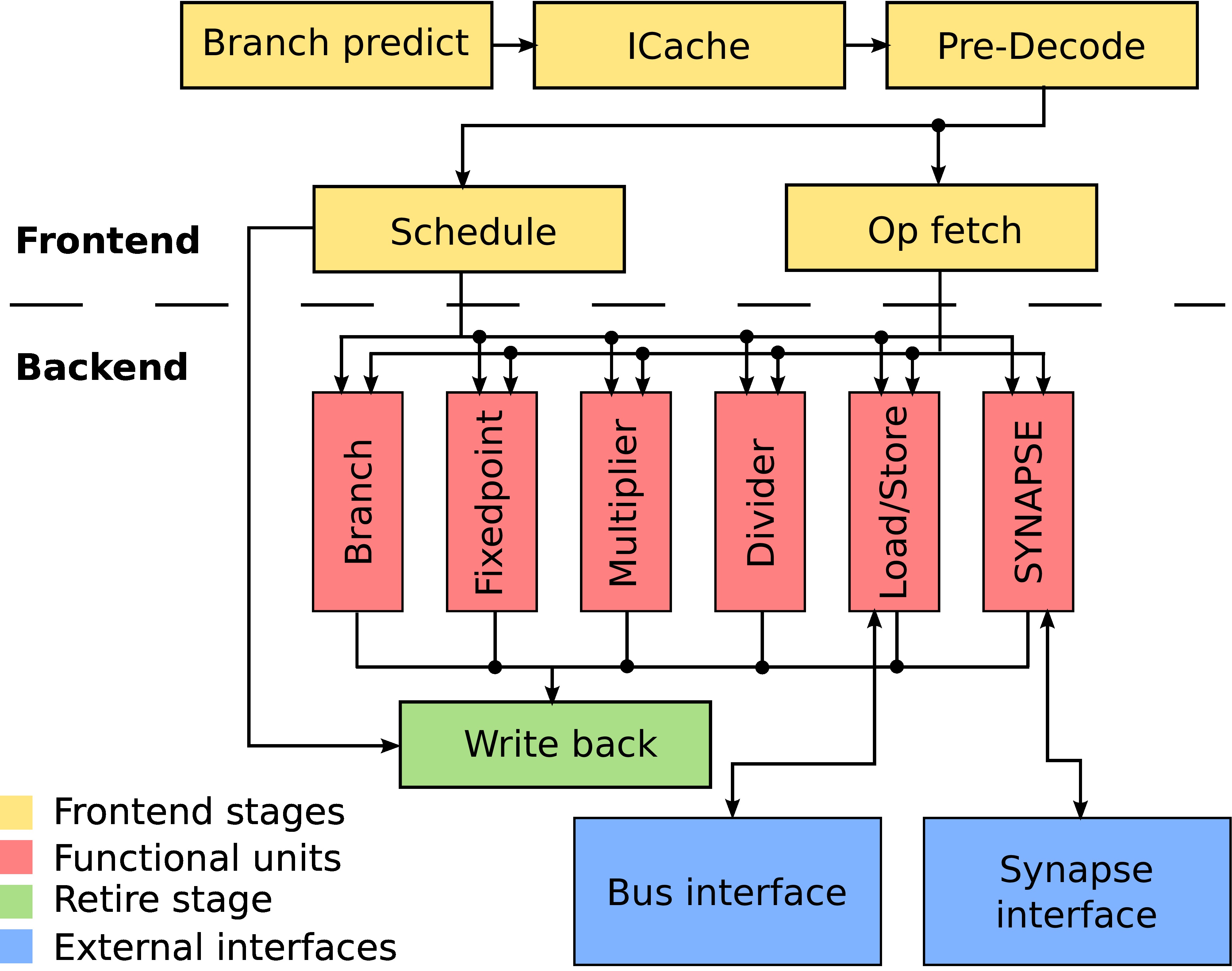}
  \caption[]{
    Micro-architecture of the embedded plasticity processor.
    The design is separated into frontend and backend.
    The frontend takes four clock cycles to decode instructions and issue them in-order to the applicable functional unit.
    The functional units take a minimum of two cycles.
    Writing the result back to the register file takes another cycle.
    Input/output operations are performed through a bus interface served by the load/store unit and a specialized interface to the synapse array.
  }
  \label{fig:architecture}
\end{figure}
The micro-architecture of the \emp is shown in Figure~\ref{fig:architecture}.
The frontend fetches and issues instruction in program order to the functional units.
Due to different latencies, instructions can retire out of program order to the write back stage.
For example a slow memory access may be overtaken by a quick add instruction issued after it.
Program and data are stored in a \unit[12]{kiB} memory.
A direct-mapped cache (\textit{ICache}) is used for instruction access and to avoid the von-Neumann bottleneck \citep{Backus1978}.
Branches can be predicted with a fully associative branch predictor using \unit[2]{bit} saturating counters to track branch outcome \citep[Strategy~7 in][]{Smith1998}.
The functional units include load/store for memory access, a branch facility for control transfers, fixed-point arithmetic and logical instructions including a barrel shifter, multiply and divide.
The \synapse special-function unit implements application specific instructions and registers.
It allows for accelerated weight computation and synapse access.

An important goal for our proposed design is to maintain small area requirements to allow integration into the existing \waferscale.
To this end, we chose in-order issue of instructions to avoid additional control logic associated with tracking of instructions and reordering.
However, out-of-order completion can be achieved with relatively small area overhead using a result shift-register \citep{smith1985implementation} and was therefore included to improve performance.

\subsection{Model for reinforcement learning}
\label{sec:model}

To demonstrate reinforcement learning using the proposed system architecture, we chose a plasticity rule and a learning task described in~\cite{Fremaux2010}.
The R-STDP rule \citep{Izhikevich2007,Florian2007} is a three-factor synaptic plasticity learning rule that modulates classical two-factor STDP with a reward-based success signal $S$.
At the end of each trial of the learning task, a reward $R$ is calculated according to the performance of the network and is used to modify the weights according to the learning rule.

\subsubsection{Network model}
\label{sec:networkmodel}
The network we simulate consists of two layers, connected with plastic synapses using the reward-modulated learning rule. The input layer consists of units repeating a given set of spike trains. The output layer consists of spiking neurons, being excited by the fixed activity from the input layer.

The original network in~\cite{Fremaux2010} uses the simplified Spike Response Model \citep[SRM$_\textrm{0}$,][]{gerstner02book} for the output neurons.
It is an intrinsically stochastic neuron that emits spikes based on the exponentially weighted distance to the threshold.
In hardware the most commonly used neuron type is the deterministic leaky integrate-and-fire (LIF).
The proposed system would use the hardware neuron reported in \cite{millner10} that can be operated as Adaptive Exponential Integrate-and-Fire \citep[AdEx,][]{brette_05} or conventional LIF model.
Since a certain amount of randomness in the firing behavior is required for reinforcement learning, we add background noise stimulation in the form of Poisson processes.
%

A tabular description of the network model can be found in Table~\ref{tab:model}.
$N_U$ input units project onto $N_{T}$ neurons that are additionally stimulated by $N_B$ random background sources.
All neurons are connected to all inputs, but each has individual random stimulation from equally sized and disjoint subsets of the random background.
In every trial the same input spike pattern is presented, but the background noise realization is different.

For each input $i=0\ldots N_U-1$, the input pattern consists of randomly drawn spike times $S_{ij} \in \mathcal{U}\left( 0, t_\text{trial} \right)$ with $j=0\ldots N_\text{stim}-1$, where $\mathcal{U}\left( 0, t_\text{trial} \right)$ is  the uniform distribution on the interval $\left[ 0, t_\text{trial} \right]$.
All simulations use the same input spike times $S_{ij}$ that are generated once to ensure comparability.

Weights for the random background have a uniform value $w_B$, so that every background spike causes the neuron to fire.
Weights for input synapses are initialized to $w_S$, chosen so that single input spikes do not cause firing.
See Table~\ref{tab:modelnum} for the numerical values.

\begin{table*}
\centering\small
\caption{
  Description of the network model used for the learning task after \cite{Nordlie2009}.
  See Table~\ref{tab:modelnum} for numerical values of the parameters.
}
\label{tab:model}
\begin{tabular}{p{4cm}p{4cm}p{8cm}}
\toprule
\multicolumn{3}{l}{\textbf{A: Model summary}} \\
\midrule
Populations             & \multicolumn{2}{p{12cm}}{Three: input $U$, random background $B$, target $T$} \\
Connectivity            & \multicolumn{2}{p{12cm}}{Feed-forward} \\
Neuron model            & \multicolumn{2}{p{12cm}}{Leaky-integrate-and-fire, fixed voltage threshold, fixed absolute refractory period (voltage clamp)} \\
Synapse model           & \multicolumn{2}{p{12cm}}{Exponentially shaped post-synaptic conductances} \\
Plasticity              & \multicolumn{2}{p{12cm}}{Three-factor STDP} \\
Input                   & \multicolumn{2}{p{12cm}}{Fixed-length spike-trains with uniformly distributed firing times} \\
\midrule
\multicolumn{3}{l}{\textbf{B: Populations}} \\
\midrule
Name                    & Elements                  & Population size \\
$U$                     & Stimulus generator        & $N_U$ \\
$B$                     & Poisson generator         & $N_B$ \\
$T$                     & LIF neurons               & $N_T$ \\
\midrule
\multicolumn{3}{l}{\textbf{C: Connectivity}} \\
\midrule
Source                  & Target                    & Pattern \\
$U$                     & $T$                       & All-to-all, initial weights $w_S$ \\
$B$                     & $T$                       & Non-overlapping 250 $\rightarrow$ 1, weight $w_B$ \\
\midrule
\multicolumn{3}{l}{\textbf{D: Neuron and synapse model}} \\
\midrule
Name                    & LIF neuron                & \\
Type                    & \multicolumn{2}{p{10cm}}{Leaky integrate-and-fire, exponential-shaped synaptic conductances} \\
Sub-threshold dynamics  & \multicolumn{2}{p{10cm}}{
  $\begin{cases}
    C_m \frac{dV}{dt} = g_L\left(E_L-V\right)+g(t)\left(E_e-V\right) & \text{if}\; t > t^*+\tau_\text{ref} \\
    V(t) = V_\text{reset} & \text{else} \\
  \end{cases}$
} \\
                        & \multicolumn{2}{p{10cm}}{$g(t)=w \exp\left( -t/\tau_\text{syn} \right)$} \\
Spiking                 & \multicolumn{2}{p{10cm}}{$\text{if}\; V(t-) < V_\text{th} \wedge V(t+) \ge V_\text{th}$} \\
                        & \multicolumn{2}{p{10cm}}{\hspace{5mm}1. set $t^* = t$} \\
                        & \multicolumn{2}{p{10cm}}{\hspace{5mm}2. emit spike with time-stamp $t^*$} \\
\midrule
\multicolumn{3}{l}{\textbf{E: Plasticity}} \\
\midrule
Name                    & Three-factor STDP& \\
Spike pairing scheme    & \multicolumn{2}{p{10cm}}{Reduced symmetric nearest-neighbor \citep{morrison08_stdp}} \\
Weight dynamics         & \multicolumn{2}{p{10cm}}{$\Delta = Sa(t)$} \\
                        & \multicolumn{2}{p{10cm}}{
  $a(t) = \sum_{\substack{i\\t_i<t}} A_\pm\exp\left(\frac{|\Delta t_i|}{\tau_\pm}\right) \exp\left(-\frac{t-t_i}{\tau_e}\right)$
} \\
                        & \multicolumn{2}{p{10cm}}{$w \in [w_\text{min}, w_\text{max}]$} \\
\midrule
\multicolumn{3}{l}{\textbf{F: Input}} \\
\midrule
Type                    & Target                    & Description \\
Stimulus generator      & $U$                       & $N_\text{stim}$ spikes at random firing times distributed uniformly within the trial duration. \\
Poisson generators      & $B$                       & Independent Poisson spike-trains with rate $\nu_B$ \\
\bottomrule
\end{tabular}
\end{table*}

\begin{table}
  \centering
  \caption[]{
    Numerical values for parameters.
    For parameter definitions see Table~\ref{tab:model} and text.
  }
  \label{tab:modelnum}
  \begin{tabular}{l l}
  \toprule
  Parameter           & Value \\
  \midrule
  $N_U$               & $250$ \\
  $N_B$               & $N_T\cdot 250$ \\
  $N_T$               & $5$ \\
  $C_m$               & $\unit[500]{pF}$ \\
  $g_L$               & $\unit[10]{nS}$ \\
  $E_L$               & $\unit[-70]{mV}$ \\
  $E_e$               & $\unit[0]{mV}$ \\
  $\tau_\text{ref}$   & $\unit[10]{ms}$ \\
  $V_\text{reset}$    & $\unit[-60]{mV}$ \\
  $V_\text{th}$       & $\unit[-50]{mV}$ \\
  $A_\pm$             & $\unit[\pm 32]{pS}$ \\
  $\tau_\pm$          & $\unit[20]{ms}$ \\
  $\tau_e$            & $\unit[0.1\ldots1000]{s}$ \\
  $w_\text{min}$      & $\unit[0]{nS}$ \\
  $w_\text{max}$      & $\unit[0.5]{nS}$ \\
  $w_B$               & $\unit[20.0]{nS}$ \\
  $w_S$               & $\unit[0.21]{nS}$ \\
  $\hat{W}$           & $\unit[0.45]{nS}$ \\
  $\nu_B$             & $\unit[0.008]{Hz}$ \\
  $t_\text{trial}$      & $\unit[1]{s}$ \\
  \bottomrule
  \end{tabular}
\end{table}
\subsubsection{Synaptic plasticity model}
In the reward modulated STDP learning rule, the outcome of standard STDP drives so-called eligibility trace changes $\Delta e_k$:
\begin{eqnarray}
  \Delta e_k = \eta A_\pm\exp\left(-\frac{\lvert\Delta t_k\rvert}{\tau_\pm}\right) \, ,
\label{eqn:stdp}
\end{eqnarray}
with learning rate $\eta$, time-difference between pre- and post-synaptic spike $\Delta t_k$ for the $k\text{-th}$ pair, STDP time constant $\tau_+$ for pre-before-post pairings, $\tau_-$ for post-before-pre pairings, and, in the same fashion, amplitude parameters $A_\pm$.
The $\Delta e_k$ are accumulated on a per-synapse eligibility trace $e$.
This trace decays exponentially according to:
\begin{eqnarray}
  e(t) = \sum_{\substack{k\\t_k<t}} \Delta e_k \exp\left(-\frac{t-t_k}{\tau_e}\right)
\label{eqn:eligibility}
\end{eqnarray}
with time-constant $\tau_e$ of the decay and $t_k$ being the time of the post-synaptic spike for pre-before-post pairings and of the pre-synaptic spike otherwise.

To calculate the weight update, a success signal $S$ is used as modulating third factor.
It represents the difference between reward received $R$ and a running average of reward $\overline{R}$
\begin{equation}
  S = R - \overline{R}
\label{eqn:success}
\end{equation}
The reward is given at the end of each trial according to the learning task as defined in the next section.
The running average is calculated as $\overline{R}_{n+1} = R_n + \left(R_n - \overline{R_n}\right)/5$ for the $n\text{-th}$ trial.
The weight update is then given by
\begin{eqnarray}
  \Delta = S e\left(t_\text{trial}\right)
\label{eqn:weightup}
\end{eqnarray}
with the trial duration $t_\text{trial}$.

In \cite{Fremaux2010} different time constants for pre-before-post ($\tau_+ = \unit[20]{ms}$) and post-before-pre ($\tau_- = \unit[40]{ms}$) are used.
The amplitudes $A_+$ and $A_-$ are chosen so that both parts are balanced, i.e. $A_+\tau_+ = -A_-\tau_-$.
Synapses of the \waferscale are designed for time constants of \unit[20]{ms}.
We do not want to assume, that this can be increased by a factor of two and therefore, we reduce $\tau_-$ to the same value as $\tau_+$.
Consequently we also use identical amplitudes to keep the STDP window $W$ balanced.
The plasticity rule described in this section represents the theoretical ideal model for our comparison that we refer to as the baseline model.
Section~\ref{sec:hwconstr} describes how this is mapped to hardware and the resulting constraints.

\subsubsection{Learning task}
\label{sec:learningtask}
In reinforcement learning, reward given is determined by the nature of the learning task considered. In our case, the goal of the network is to reproduce a given target spike train.
Hence, reward should be given in proportion to the similarity of the actual and target outputs, as measured by some metric.
Here, we use a normalized version of the metric $D^\text{spike}[q]$ by \cite{victor1996}. 
$D^\text{spike}[q]$ represents the minimal cost of transforming the output of a trial into the target pattern by adding, deleting and shifting spikes.
Adding and deleting have unit cost, while shifting by $\Delta t$ has a cost of $q\Delta t$.
For $\Delta t > 2/q$, deleting the spike and adding a new one at the correct time is cheaper than shifting it.
Therefore, the parameter $q$ controls the precision of the comparison.
The cost parameter is set to $1/q = \unit[20]{ms}$ for our simulations.

Thus in a trial where neuron $j$ fires with a spike train $X_\text{out,j}$ and the target was $X_\text{target}$, the contribution of neuron $j$ to the reward is
\begin{equation}
  R_j = 1- \frac{D^\text{spike}[q]\left( X_\text{out,j}, X_\text{target} \right)}{N_\text{out,j} + N_\text{target}} \, ,
  \label{eqn:reward}
\end{equation}
where $N_\text{out,j}$ and $N_\text{target}$ are the number of spikes in $X_\text{out,j}$ and $X_\text{target}$ respectively.
Because $D^\text{spike}[q]$ is bound to $\left[ 0, N_\text{out,j} + N_\text{target} \right]$, $R_j$ is limited to $\left[ 0, 1 \right]$.
The total reward $R$ used for the weight update is the average of $R_j$ over all $N_T$ neurons.

The target spike train is generated by simulating the neural network with a set of 
reference weights $W_{ij}$ 
for inputs $i=0\dots N_U-1$ and neurons $j=0\dots N_T-1$.
All simulations use the same set of 
reference weights 
 to ensure fair comparison:
\begin{eqnarray}
  W_{ij} =
  \begin{cases}
    \hat{W} \sin\left(\frac{i\pi}{N_U}\right)   & \text{if}\: 0 \le i \le \frac{N_U}{2} \\
    0                                           & \text{if}\: \frac{N_U}{2} < i < N_U
  \end{cases}
  \label{eqn:refweights}
\end{eqnarray}
with $\hat{W} = \unit[0.45]{nS}$.
An example of an output spike pattern produced by the network is shown in Figure~\ref{fig:raster}.
A new target spike train is generated at the beginning of every simulation run.
Its firing times can be different even for identical weights and stimulation, because of the random background stimulation.

\subsubsection{Simulated hardware constraints}
\label{sec:hwconstr}
The baseline plasticity model described in Eqs~\ref{eqn:stdp}-\ref{eqn:weightup} can not be reproduced exactly by the proposed system.
This results in two distinct classes of effects:
trade-offs introduced on purpose to reduce costs, for example in area, and non-ideal behavior of the hardware system.

In the first category, we analyze the effect of discretized weights and a limited access to analog variables by software running on the \emp.
For the second category we study leakage in analog circuits and timing effects caused by finite processor speed and communication latencies.

\paragraph{Discrete weights}
In the hardware system, synaptic weights are discretized since they are stored as digital values in the synapse circuit.
The number of bits per synapse is a critical design decision when building a neuromorphic hardware system.
Having fewer bits saves wafer area, so that more synapses can be implemented.
More bits, on the other hand, allow for a higher dynamic range of the synaptic efficacies.
The weight resolution also defines the minimum step size that can be taken by a learning rule.
To analyze the sensitivity of learning performance to weight resolution, we modify the baseline model to use discrete weights with different numbers of bits.
On a learning rule update, we precisely calculate the new weight (\unit[64]{bit} floating point) and round it to the nearest representable discrete weight value.
The tie-breaking rule is round-to-even.

In the case of non-continuous weights with \unit[$r$]{bits}, all updates with 
\begin{equation}
  |\Delta| < \frac{1}{2} \frac{w_\text{max} - w_\text{min}}{2^{r} -1}
  \label{eqn:uploss}
\end{equation}
are discarded by rounding.
Here $w_\text{min}$ and $w_\text{max}$ are the minimum and maximum weight values that can be represented and $\Delta$ is the true weight update (see Equation~\ref{eqn:weightup}).
Fewer bits per synapse means that more updates are discarded, causing the effective learning rule to increasingly deviate from the baseline learning rule.

A workaround to this problem is to perform discretized updates $\Delta_d$ probabilistically, depending on the exact weight update $\Delta$ as given by Equation~\ref{eqn:weightup}. In this way, some of the updates that would otherwise be lost can be preserved. Using the correct update probabilities results in the average weight change being identical to that of the baseline model, i.e., without discretization.

To see this, we note that $\Delta_d$ can only assume values that are multiples of the discretization step $\delta_r = \left(w_\text{max}-w_\text{min}\right)/\left(2^r-1\right)$, assuming $w_\text{min} = 0$.
If the baseline weight change $\Delta$ is between the $k\text{-th}$ and $(k-1)\text{-th}$ step,  the discrete update $\Delta_d$ is picked from those with probability $p = \Pr\left(\Delta_d = k\delta_r \mid \Delta\right)$ and $1-p$, respectively.
Such a scheme leads to the average update $\left< \Delta_d \right>$ for a given $\Delta$ being 
\begin{eqnarray}
  \left< \Delta_d \right> &=& k\delta_r p + (k-1)\delta_r(1-p) \\
                           &=& \delta_r(k-1)+\delta_r p \, .
  \label{eqn:avgprobup}
\end{eqnarray}
By picking $p$ as
\begin{eqnarray}
  p = \frac{\Delta - (k-1)\delta_r}{\delta_r} \, ,
  \label{eqn:probup}
\end{eqnarray}
it holds that $\left< \Delta_d \right> = \Delta$.

\paragraph{Baseline model with added noise}
When performing weight updates probabilistically, randomization introduces additional noise to the weight dynamics.
This noise is not present in the baseline model with continuous weights.
Therefore, adding an equivalent amount of random noise to the baseline simulation allows for a more accurate assessment of weight discretization with probabilistic updates.

With every update, probabilistic rounding introduces an error $z = \Delta_d - \Delta$.
For simplification, we introduce $\epsilon \in [0, \delta_r)$ and substitute $\Delta = (k-1)\delta_r + \epsilon$ in Equation~\ref{eqn:probup} to get $p = \epsilon / \delta_r$.
Then, $z$ is distributed according to 
\begin{eqnarray}
 \Pr\left( z \mid \epsilon \right) &=&
 \begin{cases}
   p    & \text{if}\,\, z = \delta_r - \epsilon \\
   1-p  & \text{if}\,\, z = -\epsilon \\
   0    & \text{otherwise.}
 \end{cases}
 \label{eqn:pzcond}
\end{eqnarray}
We are now interested in the unconditional probability distribution $\Pr(z)$ to add noise shaped accordingly to the baseline simulation with continuous weights.
This is given by
\begin{eqnarray}
  \Pr(z) &=& \int_0^{\delta_r} \Pr(z\mid \epsilon) \Pr(\epsilon) \mathrm{d}\epsilon.
  \label{eqn:pz}
\end{eqnarray}
Assuming $\epsilon$ to be uniformly distributed in its allowed interval gives $\Pr(\epsilon) = \delta_r^{-1}$.
Using the Kronecker-Delta $\delta$ to write down $\Pr(z\mid\epsilon)$ with $p = \epsilon/\delta_r$ (Equation~\ref{eqn:probup}) gives:
\begin{eqnarray}
  \Pr(z) &=& \frac{1}{\delta_r} \int_0^{\delta_r} \frac{\epsilon}{\delta_r} \delta\left( z-\delta_r+\epsilon \right) + \left( 1 - \frac{\epsilon}{\delta_r} \right) \delta\left( z+\epsilon \right) \mathrm{d}\epsilon \\
         &=&
  \begin{cases}
    \frac{\delta_r-z}{\delta_r^2}   & \text{for}\,\, 0 < z < \delta_r \\
    \frac{\delta_r + z}{\delta_r^2} & \text{for}\,\, -\delta_r < z \le 0 \\
  \end{cases}
  \label{eqn:pz2}
\end{eqnarray}
Equation~\ref{eqn:pz2} describes a triangular shaped probability density for the noise introduced by probabilistic updates.
As is to be expected, the noise is bounded by $\pm\delta_r$.

\paragraph{Thresholded readout}
The eligibility trace is implemented using the analog accumulation in the synapse unit.
For every spike pair, Equation~\ref{eqn:hwdeltae} is evaluated and the corresponding eligibility trace change is added as charge on the local storage capacitors $a_+$ and $a_-$ respectively.
These values are not directly accessible to the \emp. Instead, using the evaluation unit described in Section~\ref{sec:implementing} with threshold $\Theta = a_\text{th} - a_\text{tl}$, accumulation trace $a = a_+ - a_-$, configuration bits $e_{ac}^+=1$, $e_{aa}^+=1$, $e_{ca}^+=0$, $e_{cc}^+=0$ for the evaluation of $b_+$ and $e_{ac}^-=0$, $e_{aa}^-=0$, $e_{ca}^-=1$, $e_{cc}^-=1$ for $b_-$, the readout computes
\begin{eqnarray}
  b_\pm =
  \begin{cases}
    1 & \text{if } \pm\left(a_+ - a_-\right) > \Theta \\
    0 & \text{otherwise}
  \end{cases} \, .
  \label{eqn:diffreadout}
\end{eqnarray}
The weight update with threshold readout $\Delta_t$ is then performed using an update constant $A$
\begin{equation}
  \Delta_t = SA \left( b_+ - b_- \right) \, .
  \label{eqn:readoutweightup}
\end{equation}

The paraameters $\Theta$ and $A$ should be chosen so as to minimize the deviation introduced by calculating weights according to Equation~\ref{eqn:readoutweightup} instead of Equation~\ref{eqn:weightup}. Ideally, one would like to satisfy $\left< \Delta_t \right> = \Delta$. However, detailed analysis of the simulations (not shown) showed that the eligibility trace distributions for different synapses at different stages of learning were very different. In that context, choosing parameters $\Theta$ and $A$ that minimize the difference between the baseline change $\Delta$ and the average effective change $\left< \Delta_t \right>$ for a particular synapse would not in general have the same effect for other synapses. Instead, we resort to a heuristic method to fix global threshold and update constant, described below, and assess its effectiveness in simulations.

For the simulations presented here, a precursor run over 100 trials without learning was used to measure the final absolute eligibility value $\left< |a| \right>$ averaged over all readout operations.
The threshold $\Theta$ was then set to $\Theta^* = \left< |a| \right>$ for the actual learning  simulation.
In this way, the average (across synapses) final eligibility value encountered during weight updates is close to the threshold.
This represents a trade-off between exceeding the threshold only seldom, but then causing large -- possibly disruptive -- weight changes, and exceeding the threshold often, but only applying small changes.

With $N_p\left(\Theta\right)$ being the number of readout operations that exceed the threshold, i.e. $b_+$ or $b_-$ are non-zero, and the total number of readout operations $N$, the update constant $A$ is set to
\begin{equation}
  A^* = \frac{N}{N_p(\theta^*)} \theta^* \, .
  \label{eqn:opta}
\end{equation}
Thereby, the mean absolute eligibility value used with the readout $N_p(\Theta^*)A^*/N$ is effectively the same as $\left< |a| \right>$ in the baseline model.

\paragraph{Analog drift}
The local accumulation units in the hardware synapses do not have a mechanism for controlled decay of the eligibility trace.
An ideal implementation of the circuit would stay unchanged over time, after a spike-pair has caused an update.
In reality there are leakage currents causing the accumulation traces $a_+$ and $a_-$ and their difference $a$ to drift.
Leakage is caused by a number of processes that depend on transistor geometry, manufacturing process, temperature and internal voltages \citep{Roy2003}.
It is therefore difficult to predict either time-scale, shape or variability of this effect.
We try to get an estimate on the sensitivity of the model to uncontrolled temporal drift, by simulating learning with a drift function $\phi_i\left( t ; a_0 \right)$.
Here $t$ is the duration of the drift and $a_0$ is the starting value for $t=0$.
The index $i$ is over all synapses and both trace polarities.
This function describes the development of $a_+\left( t \right)$ and $a_-\left( t \right)$ between spike-pair induced updates.
The accumulation value is given as the difference $a\left( t \right) = a_+\left( t \right) - a_-\left( t \right)$.
We define an exponential drift function
\begin{eqnarray}
  \phi_i\left( t ; a_0 \right) = 
  \begin{cases}
    a_0 e^{-\lambda_i t}                                               & \text{for}\,\, \lambda_i > 0 \\
    a_\text{max} - \left( a_\text{max} - a_0 \right) e^{\lambda_i t}   & \text{for}\,\, \lambda_i < 0 \\
    a_0                                                                & \text{else}\, , \\
  \end{cases} 
  \label{eqn:drift}
\end{eqnarray}
where $a_\text{max}$ is the maximum value that $a_+$ and $a_-$ can assume and $\lambda_i=1/\tau_{e,i}$ is the inverse time constant.
Positive $\lambda_i$ leads to exponential decay as it was used so far.
Negative $\lambda_i$ causes a drift away from zero, towards the limit $a_\text{max}$.
For every synapse and for positive and negative traces, $\tau_{e,i}$ is drawn from a Gaussian distribution with mean $\tau_e$ and standard deviation $m_e\tau_e$ using the mismatch factor $m_e$.
In the limit of large $t$, this allows for four final states of $a\left( t \right)$:
Decay to zero, drift to $a_\text{max}$ or $-a_\text{max}$ and remaining constant at $a_0$ for $\lambda_i = 0$.

It is important to note that we do not intend to precisely model the leakage behavior of the analog circuit.
Instead, we use a simple model capturing the essence of drifting analog values to get an estimate for the sensitivity to this effect.

\paragraph{Delayed reward}
The hardware system is a physical model of the emulated network.
Therefore, emulated time progresses continuously during network operation with the acceleration factor $\alpha$ relative to wall-clock time.
During all communication and computation, network operation continues.
The amount of reward for each trial is calculated by the control cluster, after the nominal trial duration has ended and output spike events have been transmitted to the cluster.
The success signal is then determined and sent back to the embedded processor.
Then, the plasticity program will sequentially execute the weight update for all synapses taking a certain amount of time per synapse.
This time is consumed by the synapse array access and the weight computation.

These two effects are modeled by adding a constant delay $D_R$ after the trial has finished and an update rate $\nu_s$ giving the number of updated synapses per second.
The weight update for synapse $i$ occurs at $t_i = t_\text{trial} + D_R + \frac{i}{\nu_s}$.
The order in which synapses are updated is determined by their position in the synapse array and is therefore a result of the automated mapping process.
For this study, we assume weight updates to be fast enough compared to the reward delay $D_R$ and therefore use $t_i = t_\text{trial} + D_R$.

The delay causes a deviation from the ideal model because the accumulation capacitors $a_+$, $a_-$ used to store the eligibility trace continue to decay.
The eligibility value used for the weight update is then reduced by a factor 
\begin{equation}
  \beta= \exp\left( -\frac{D_R}{\tau_e}\right)
\end{equation}
This can prevent a weight update that would have been made in the non-delayed case by reducing $a$ below the readout threshold $\Theta$.
We assume that the delay $D_R$ is known or can be estimated and lower the threshold to $\beta\Theta$.

In theory, this would allow to correct for arbitrary delay, since the exponential decay never reaches zero.
In hardware this is not the case, because the eligibility readout is subject to noise.
Therefore, after a certain delay, traces will be indiscernible from noise.
To account for this, we simulate Gaussian distributed noise $\delta a$ on the readout with standard deviation $\sigma_a$ and mean 0.
The value used for comparison to the threshold is then given by $a' = a + \delta a$.
If a signal-to-noise ratio $z^*$ is required for correct learning, a limit $D_\text{max}$ for the delay can be calculated using the signal-to-noise ratio $z(t) = a(t)/\sigma_a$
\begin{eqnarray}
  z(t) &=& \frac{a\left(t_\text{trial}\right)}{\sigma_a} \exp\left(-\frac{t-t_\text{trial}}{\tau_e}\right)
\end{eqnarray}
With $z\left(D_\text{max}+t_\text{trial}\right) = z^*$ and $a\left(t_\text{trial}\right) = a_\text{max}$, the maximally tolerable delay in the presence of noise is given by
\begin{eqnarray}
  D_\text{max} &=& -\tau_e \ln\left(\frac{z^*\sigma_a}{a_\text{max}}\right)
  \label{eqn:dmax3}
\end{eqnarray}

\subsubsection{Measuring performance}
Simulations consist of 10000 trials in 20 parallel runs with different random seeds.
At the beginning of every run, 100 trials are simulated without learning:
during this time the running average $\overline{R}$ can settle to a stable approximation of the reward.
The average over $R$ during these trials is used as the initial reward level $R_\text{before}$ of this run.
During the last 1000 trials of the simulation, it is assumed that learning has reached a stable state:
the final reward level $R_\text{after}$ is the average of $R$ over these trials.


The model is simulated using the Brian simulator \citep{goodman2008}.
Weight updates are calculated with custom Python code using the NumPy package \citep{numpy_homepage}.

\section{Results}
\label{sec:results}
In the previous section, we analyzed a synaptic learning rule~\citep{Izhikevich2007,Florian2007,Fremaux2010}, and the necessary adjustments that have to be made in order to implement it on a hardware system.
The goal of this section is to quantify the sensitivity to constraints of the system -- for example discretized weights or imperfections of analog circuits -- to identify those critical for the model.
Starting from the baseline configuration without hardware effects, we add constraints and measure their effect on the learning performance.

\subsection{Baseline}
\label{sec:baseline}
\begin{figure}
  \begin{center}
    \includegraphics[width=\figw]{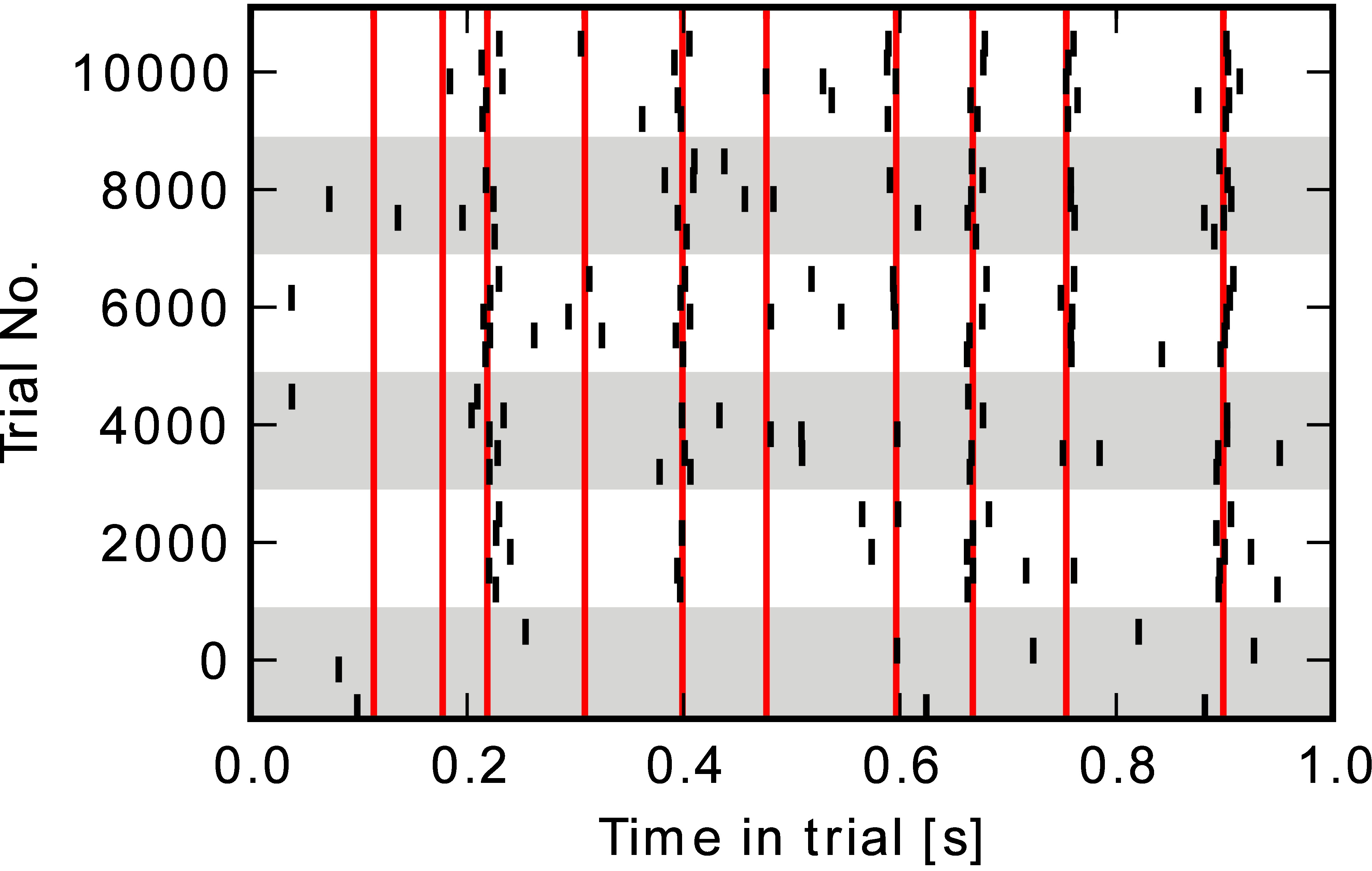}
  \end{center}
  \caption{
    Raster-plot of output spike-events for all five neurons at intervals of 2000 trials.
    Red bars indicate the target firing times.
  }
  \label{fig:raster}
\end{figure}
The baseline model implements the learning rule described in Section~\ref{sec:model} and Table~\ref{tab:model} without hardware effects, and serves as comparison for simulations including such effects.
The eligibility trace $e$ of the theoretical model is identified with the local accumulation $a$ in hardware synapses.
Thereby, changes to the weight are deferred until the success signal $S$ is given from the attached control cluster, after the produced spike train has been evaluated.
New weights are assumed to be calculated using a software program running on the \emp.

The raster plot in Figure~\ref{fig:raster} shows the output spike train at several points in time during a learning simulation.
In the beginning at trial 0, spikes are generated randomly by the background stimulation.
Later on, the network learns to produce spikes at the targeted points of time indicated with red vertical bars.
In the last trial, neurons fire close to most of the target times.
The evolution of the reward obtained in each trial averaged over 20 runs is shown in Figure~\manref{fig:discbase}{A}.
Variance in the last 1000 trials is due to the random background stimulation and to the exploratory behavior it generates in the learning rule.
Most of the performance improvement is achieved within the first 2000 trials, the final level of reward being $R_\text{after}^\text{base} = 0.54 \pm 0.05$.

\begin{figure}
  \begin{center}
    \includegraphics[width=\figw]{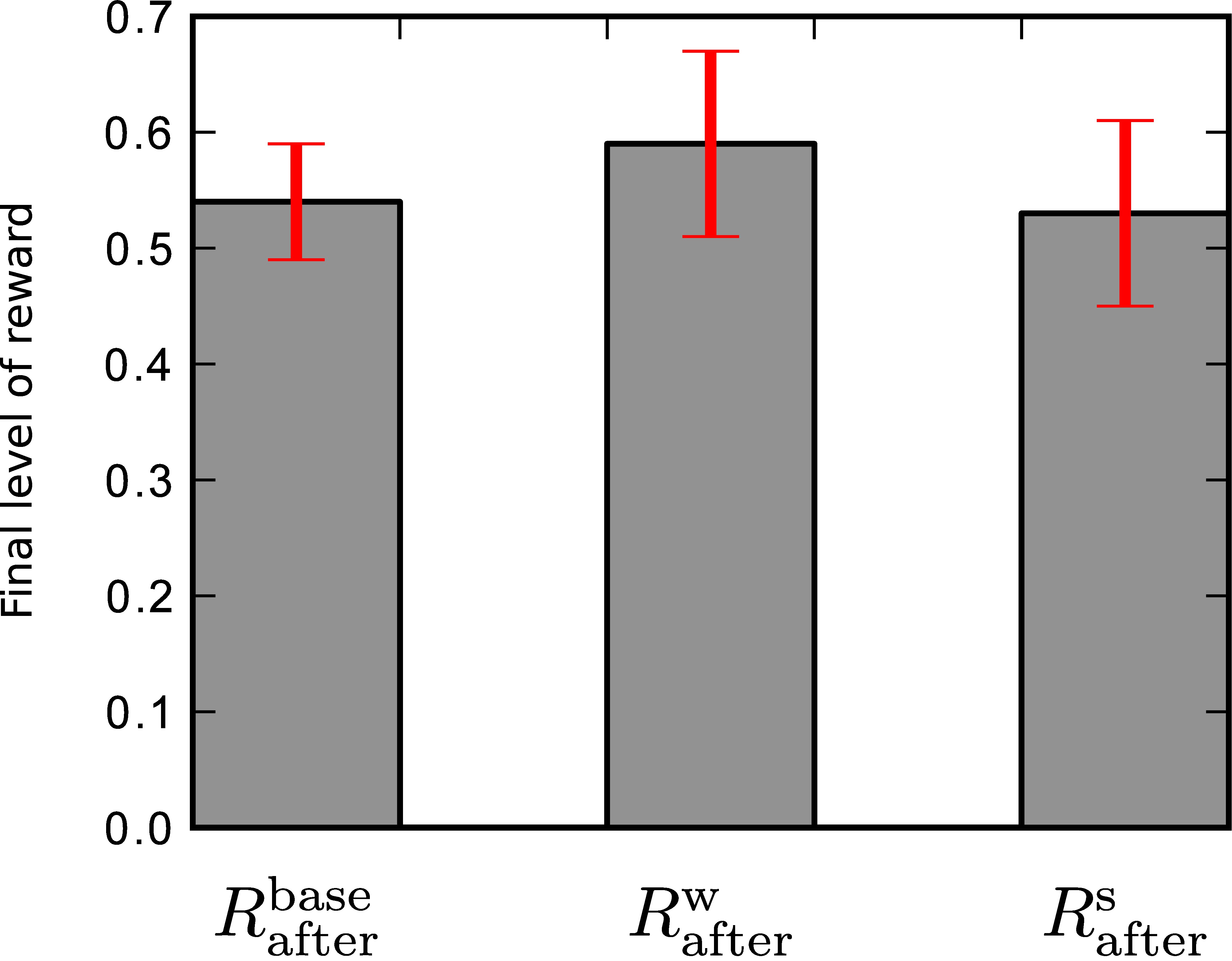}
  \end{center}
  \caption{
    Final level of reward for:
    baseline simulation, randomized reference weights, and randomized stimulation pattern.
    The final performance level of the baseline simulation $R_\text{after}^\text{base}$ using reference weights $W_{ij}$ and stimulation pattern $S_{ij}$ is comparable to the final level of reward averaged over randomly chosen reference weights $R_\text{after}^\text{w}$ and stimulation patterns $R_\text{after}^\text{s}$.
  }
  \label{fig:generalize}
\end{figure}
This is the result using one particular set of reference weights $W_{ij}$ and stimulation pattern $S_{ij}$ that were defined in Section~\ref{sec:networkmodel}.
To test how well this result generalizes to other weights and stimulation patterns we perform two additional experiments:
first of all, we randomize the reference weights, so that in 20 simulation runs the network learns with a different set of reference weights in each run.
These weights are drawn randomly from a uniform distribution, so that the $k\text{-th}$ run uses reference weights $W_{ij}^k \in \mathcal{U}\left( w_\text{min}, w_\text{max} \right)$ to generate its target spike train.
This gives a final level of reward of $R_\text{after}^{w} = 0.59\pm 0.08$ averaged over the 20 runs with different reference weights.

In the second experiment we again use the $W_{ij}$ reference weights for all 20 simulations.
The stimulation pattern is randomized by drawing new spike times for each run from a uniform distribution, so that the $k\text{-th}$ run uses spike times $S_{ij}^k \in \mathcal{U}\left( 0, t_\text{trial} \right)$ for all trials.
This gives a performance $R_\text{after}^{s} = 0.53 \pm 0.08$ averaged over the 20 different sets of stimulation patterns.


The final reward level for the baseline simulation, randomized reference weights and randomized stimulation pattern are shown in Figure~\ref{fig:generalize}.
The data show, that the from here on used special case of reference weights $W_{ij}$ and stimulation spike times $S_{ij}$ is within the performance range of randomly selected reference weights and input spike timings.
The variances on $R_\text{after}^{w}$ and $R_\text{after}^{s}$ also show that there is considerable variation in the unconstrained theoretical model.
To reduce variation in our results, so that changes caused by hardware effects are more visible, we use $W_{ij}$ and $S_{ij}$ from here on.

\subsection{Discretized weights}
\label{sec:resdisc}
In designing the neuromorphic hardware system, one is faced with a trade-off between implementing more synapses with lower bit resolution and less synapses with higher resolution.
Therefore, we would like to know how many bits are required for each synaptic weight to achieve good performance in the learning task.
We perform a three-way comparison between the baseline model, a deterministic algorithm that simply rounds calculated weights to allowed representations and a probabilistic variant as outlined in Section~\ref{sec:hwconstr}.
\begin{figure}
  \includegraphics[width=\figww]{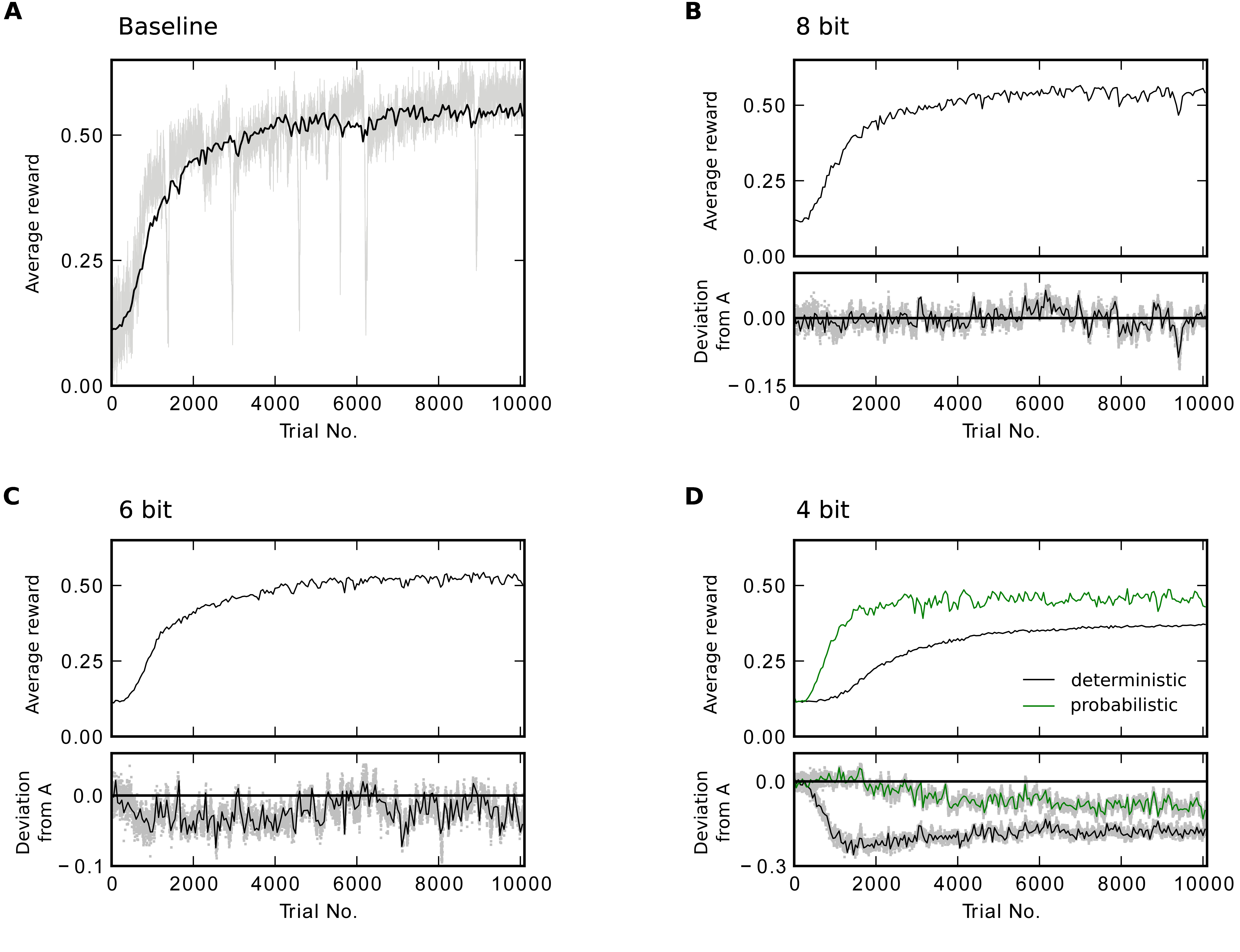}
  \caption{
    Reward traces showing the running average $\bar{R}$ (only every 50$^\text{th}$ point plotted) for different weight resolutions averaged over 20 runs.
    \mansubref{A} Baseline performance with continuous weights.
    Additionally, the light gray trace shows the reward $R$ for every trial of a single simulation.
    \mansubref{B} Performance with \unit[8]{bit} resolution.
    The lower plot shows the difference to the baseline model in \mansubref{A}.
    The shaded area shows the difference for every point in the trace instead of only for every 50$^\text{th}$.
    \mansubref{C} Performance with \unit[6]{bit} resolution.
    \mansubref{D} Performance with \unit[4]{bit} resolution.
    The black trace shows the result for deterministic updates.
    The green trace for probabilistic updates.
  }
  \label{fig:discbase}\label{fig:disc8}\label{fig:disc6}\label{fig:probweight}
  \label{fig:reward_discrete}
\end{figure}
Using deterministic weight updates, all updates satisfying Equation~\ref{eqn:uploss} do not cause a weight change.
With fewer bits more updates are lost and learning performance is expected to suffer.
This is what can be seen in Figure~\ref{fig:reward_discrete}.
The simulations shown there compare performance of the baseline model, to a constrained model with discretized weights of decreasing resolution.
Figure~\manref{fig:reward_discrete}{A} also shows the full reward trace of a single run picked arbitrarily.
The plot exhibits a number of sharp drops in reward that last for less than 15~trials, before returning to the previous performance level.
The final level of performance is not affected by these glitches.
For the \unit[8]{bit} case, performance is as good as using continuous weights (Figures~\manref{fig:discbase}{B}).
Figure~\manref{fig:discbase}{C} shows a slightly reduced performance for \unit[6]{bit}.
Using only \unit[4]{bit} with deterministic updates causes performance to degrade: it does not reach the same final level of reward (Figure~\manref{fig:probweight}{D} black trace). 
See Table~\ref{tab:performance} for the final performance values $R_\text{after}$.
Using probabilistic updates improves the performance for \unit[4]{bit} to $R_\text{after}^\text{4p} = 0.46 \pm 0.03$, which is $(85\pm 10)\%$ of the baseline level $R^\text{base}_\text{after}$ (Figure~\manref{fig:probweight}{D} green trace).

So in the task studied here, there is no gain in building synapses using more than \unit[8]{bit}.
Because weight updates are controlled by a programmable processor, it is possible to switch between deterministic and probabilistic updating even after the system has been manufactured.
In this context, a trade-off can be made between number of synapses and reachable performance by using either probabilistic \unit[4]{bit} or deterministic \unit[8]{bit} synapses.

\paragraph{Baseline with added noise}
\begin{figure}
  \begin{center}
    \includegraphics[width=\figww]{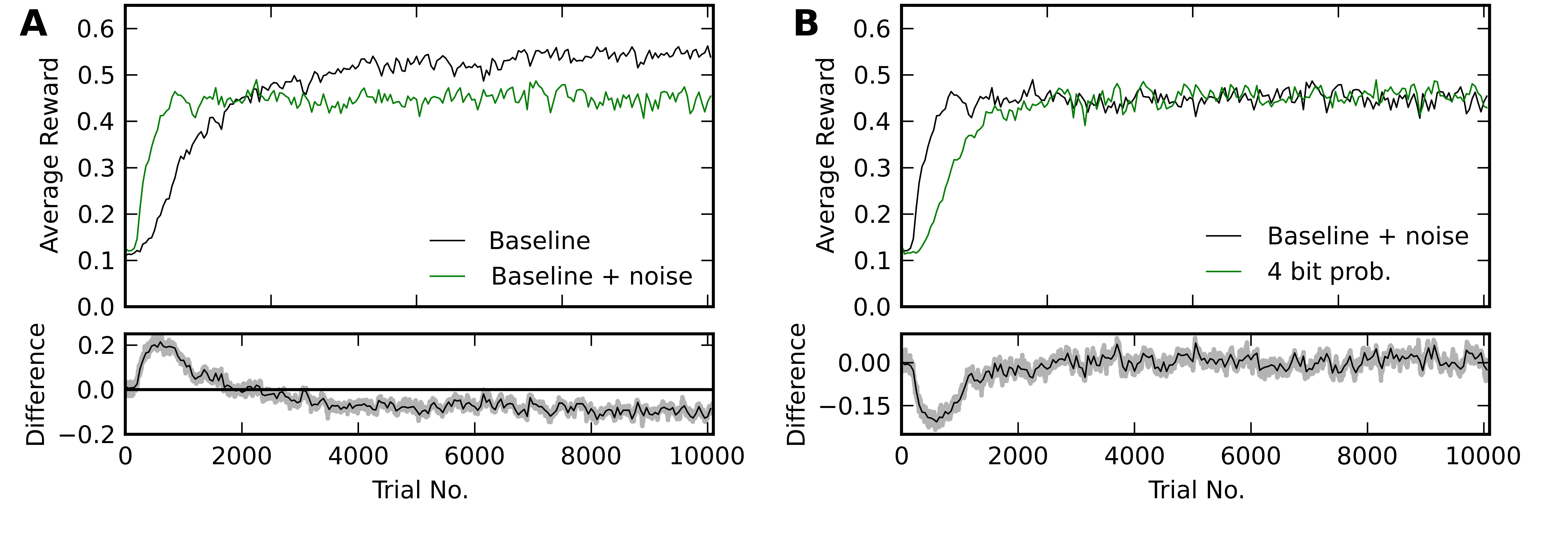}
  \end{center}
  \caption{
    \mansubref{A} Comparison of the baseline simulation with and without added noise on the weight updates.
    The lower plot shows the difference between both traces in the upper plot.
    \mansubref{B} Comparison between \unit[4]{bit} discretized weights with probabilistic updates and baseline with added noise of equivalent magnitude.
    Again, the lower plot shows the difference between both traces in the upper box.
  }
  \label{fig:probbasecmp}
\end{figure}
As discussed in Section~\ref{sec:hwconstr}, probabilistic updates introduce additional noise on the weights.
The baseline simulation with added noise uses updates $\Delta' = \Delta + z$ with $z$ drawn from the distribution given in Equation~\ref{eqn:pz2} using $r=4$.

Figure~\manref{fig:probbasecmp}{A} shows reward traces for the baseline simulation with and without added noise.
One can see, that with noise learning is initially faster, but fails to reach the same level as without.
The final level of performance in the former case is $R_\text{after}^\text{noise} = 0.45 \pm 0.03$, while it was $R_\text{after}^\text{base} = 0.54 \pm 0.05$ in the latter simulation.
Figure~\manref{fig:probbasecmp}{B} compares baseline with added noise to the case with \unit[4]{bit} weights and probabilistic updates.
Both variants reach the same final level of reward ($R_\text{after}^\text{noise} = 0.45 \pm 0.03$ and $R_\text{after}^\text{4p} = 0.46 \pm 0.03$), but with continuous weights this level is reached faster.
In conclusion, Figure~\ref{fig:probbasecmp} shows, that the achievable performance for \unit[4]{bit} resolution with probabilistic updates is limited by the added noise and not the limitation to discrete weight values.

\paragraph{Effect on weights}
\begin{figure}
  \begin{center}
    \includegraphics[width=\figww]{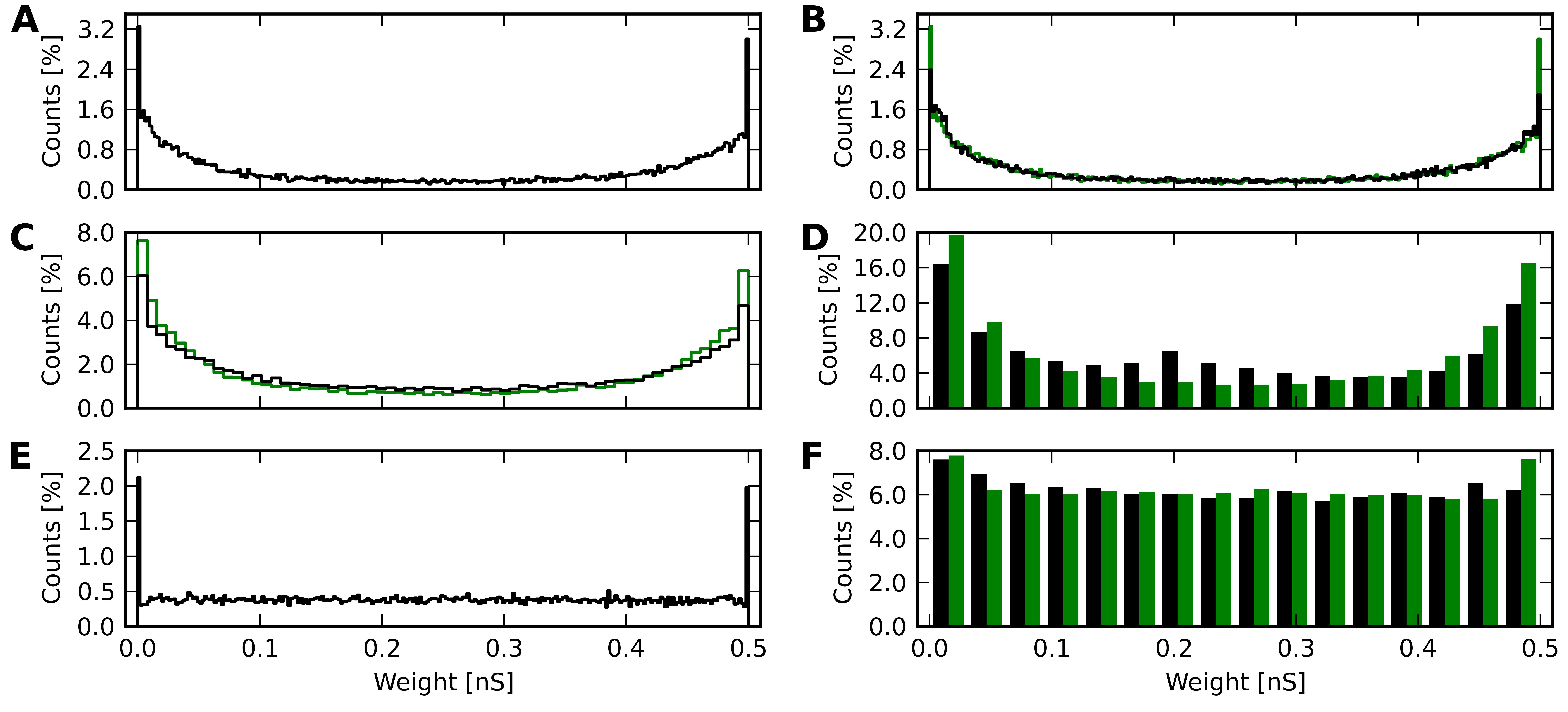}
  \end{center}
  \caption{
    Histograms of synaptic weights after learning.
    The weights from all 20~repetitions for each resolution and update mode are shown.
    \mansubref{A} Continuous weights.
    \mansubref{B} \unit[8]{bit} weights in black.
    Continuous weights are discretized to this resolution and shown in green.
    \mansubref{C} \unit[6]{bit} weights in black, again with equally binned continuous weights in green.
    \mansubref{D} \unit[4]{bit} weights with deterministic updates in black and the continuous result in green.
    \mansubref{E} Final weights for the baseline simulation with artificially added noise, of which the reward trace is shown in Figure~\ref{fig:probbasecmp}.
    \mansubref{F} Final weight histogram for \unit[4]{bit} resolution with probabilistic updates in black.
    Now the green bars give the distribution of weights from the baseline simulation with added noise.
  }
  \label{fig:weightdistdisc}
\end{figure}
Besides comparing the received reward, it is also informative to compare the distribution of synaptic weights after learning for the different weight resolutions.
Figure~\ref{fig:weightdistdisc} shows histograms of weights for different resolutions and deterministic and probabilistic updating.
The weights of the baseline simulation are given in Figure~\manref{fig:weightdistdisc}{A} and with added noise for $r=4$ in Figure~\manref{fig:weightdistdisc}{E}.
For discretized weights with deterministic updates, the distribution from Figure~\manref{fig:weightdistdisc}{A} binned to the respective resolution is also shown in green (Figure~\manref{fig:weightdistdisc}{B-D}).
For Figure~\manref{fig:weightdistdisc}{F}, the green bars show the weights from the baseline simulation with added noise binned to \unit[4]{bit}.

The baseline histograms (Figure~\manref{fig:weightdistdisc}{A,E}) are bimodal with peaks at the maximum and minimum allowed weights.
This is also the result, one would get for an unsupervised additive STDP rule \citep{morrison08_stdp}.
With discretized weights and deterministic updates, the bi-modality is maintained.
For \unit[6 and 4]{bit} an increasing deviation from the rounded baseline histogram is apparent.
Here, more weights lie in the central region, so that the counts are lower than baseline towards the minimum and maximum weights.
For \unit[4]{bit} with deterministic updates (Figure~\manref{fig:weightdistdisc}{D}) a local maximum at \unit[0.2]{nS} can be observed.
This corresponds to the initial weight $w_S = \unit[0.21]{nS}$ and indicates that many synapses have not been updated at all or only with small increments.

The results of a Kolmogorov-Smirnov (KS) test between the baseline distribution shown in Figure~\manref{fig:weightdistdisc}{A} and the respective result obtained with discrete weights is shown in Table~\ref{tab:performance}.
The baseline distribution was rounded to the weight resolution of the respective simulation for the test.
The data show increasing deviation with smaller weight resolution.
The obtained p-values indicate, that the distributions are not identical to the discretized baseline case ($p = 0.35$ for \unit[8]{bit} and $p<0.01$ otherwise).
Note, that the distribution is also different from the continuous baseline distribution, since it is discrete.

The root-mean-square error of the weights as compared to the baseline simulation is given by
\begin{eqnarray}
  E_w &=& \sqrt{ \frac{1}{N_UN_T} \sum_{i=0}^{N_UN_T} \left( w_i - \left< w^\text{base}_i \right> \right)^2 }.
  \label{eqn:rmse}
\end{eqnarray}
Here, $\left< w_i^\text{base} \right>$ is the $i\text{-th}$ weight averaged over 20 repetitions of the baseline simulation.
Averaged over the individual runs of the baseline simulation itself, this gives $\left<E_w^\text{base}\right> = \unit[(0.10\pm 0.06)]{nS}$.
For \unit[8, 6 and 4]{bit}, this increases to $\left<E_w^8\right> = \unit[(0.11\pm 0.06)]{nS}$, $\left<E_w^6\right> = \unit[(0.12\pm 0.06)]{nS}$, and $\left<E_w^4\right> = \unit[(0.17\pm 0.07)]{nS}$.
Compared to the total weight range of only \unit[0.5]{nS}, those are large deviations.
Since already the baseline simulation shows a root-mean-square error of 20\% of this range, it can be concluded, that learning does not produce a single fixed set of weights.
This is either due to redundancy in the weights or irrelevant synapses.

When noise on weight updates is added to the simulation, the distribution of final weights changes (Figure~\manref{fig:weightdistdisc}{E}).
Here, the histogram is still bimodal with peaks at the weight boundaries, but in-between the distribution is flat.
The weight noise modifies weights by up to $\delta_r \approx \unit[0.03]{nS}$ in each update (see Equation~\ref{eqn:pz2}).
This acts as a diffusion process smoothing the weight distribution.
For \unit[4]{bit} weights with probabilistic updates (Figure~\manref{fig:weightdistdisc}{F}), the histogram is also flattened compared to the variant with deterministic updates (Figure~\manref{fig:weightdistdisc}{D}).
The result is qualitatively in good agreement with the rounded weights from the baseline simulation with added noise.
The root-mean-square error using weights from the baseline simulation with added noise as reference is $\left<E_w^\text{4p}\right> = \unit[(0.15\pm 0.03)]{nS}$.
The KS test reveals a smaller deviation from the baseline simulation with noise compared to the \unit[4]{bit} case with deterministic updates (No.~6 compared to no.~5 in Table~\ref{tab:performance}).
However, the test also shows the weight distributions to not be identical ($p < 0.01$).

\subsection{Thresholded readout}
\label{sec:eligthresh}
The hybrid approach of combining processor based digital computing with analog spe\-cial-func\-tion units necessitates an interface between these two.
At this interface some form of analog-to-digital conversion (ADC) has to take place.
The simplest form of ADC is comparison to a threshold.
We next ask whether such a simple interface is sufficient for good performance on the learning task.
\begin{figure}
  \includegraphics[width=\figww]{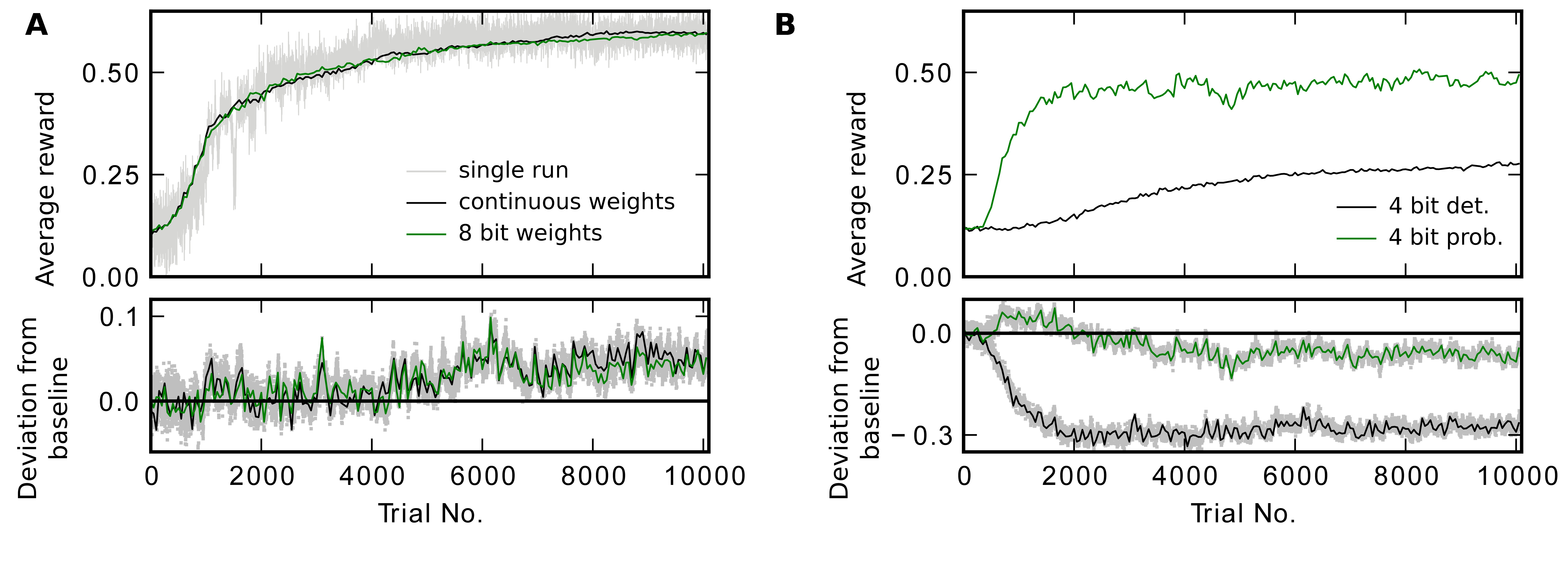}
  \caption{
    Performance with threshold readout.
    As in Figure~\ref{fig:reward_discrete} the running average of the reward $\bar{R}$ is plotted averaged over 20 runs.
    The lower plots show the difference to the baseline trace in Figure~\manref{fig:reward_discrete}{A}.
    \mansubref{A} Performance traces for continuous and \unit[8]{bit} weights.
    In gray reward $R$ for every trial in a single run with continuous weights is shown.
    \mansubref{B} Performance traces for \unit[4]{bit} resolution with deterministic and probabilistic updates.
  }
  \label{fig:discrete_elig_thresh}
\end{figure}
Figure~\ref{fig:discrete_elig_thresh}
shows performance for different weight resolutions compared to baseline using the thresholded readout.
In contrast to the simulations shown in Figure~\ref{fig:reward_discrete}, updates are now calculated according to Equation~\ref{eqn:readoutweightup} instead of Equation~\ref{eqn:weightup}.
In particular, Equation~\ref{eqn:readoutweightup} does not directly use the eligibility trace $e(t_\text{trial})$, but the evaluation bits $b_+, b_-$ determined by the readout mechanism (Equation~\ref{eqn:diffreadout}).
Performance in the case of continuous, \unit[8 and 6]{bit} synapses (\unit[6]{bit} with threshold readout mechanism not shown) qualitatively shows the same picture with and without threshold readout (compare Figures~\ref{fig:reward_discrete} and~\ref{fig:discrete_elig_thresh}):
Resolutions of \unit[8 and 6]{bit} reach good performance while \unit[4]{bit} with deterministic updates is degraded.
The precise values of the final reward $R_\text{after}$ given in Table~\ref{tab:performance} indicate a small improvement of $0.06\pm 0.04$ in reward by the threshold mechanism for \unit[8 and 6]{bit}.
When comparing traces for weights of the same resolution in Figures~\ref{fig:reward_discrete} and~\ref{fig:discrete_elig_thresh}, those with threshold readout (Figure~\ref{fig:discrete_elig_thresh}) show less variability between trials.
For example, the trace of the single run in Figure~\manref{fig:reward_discrete}{A} exhibits more noise than the one in Figure~\manref{fig:discrete_elig_thresh}{A}.
The variability can be quantified by the standard deviation $\sigma_S$ of the success signal $S$ (see Equation~\ref{eqn:success}).
For a resolution of \unit[8]{bit}, $\sigma_S = 4.0\times 10^{-5}$ is reduced to $\sigma_S = 1.2\times 10^{-5}$, when using the threshold readout.
This is caused by the smoothing effect of the readout threshold, which effectively replaces extreme values of the eligibility trace $e(t_\text{trial})$ with the update constant $A=A^*$.
The update constant $A^*$ is determined heuristically according to Equation~\ref{eqn:opta}.

When using probabilistic updates (Figure~\manref{fig:discrete_elig_thresh}{B}, green trace), the performance level of the baseline simulation with added noise on the weights of equivalent magnitude is also slightly surpassed (see Nos.~2 and~9 in Table~\ref{tab:performance}).
With deterministic updates and \unit[4]{bit} synapses, performance is further reduced by $0.10\pm0.05$ using the threshold readout (black traces in Figures~\manref{fig:reward_discrete}{D} and~\manref{fig:discrete_elig_thresh}{B}).

Hence the simple readout method consisting in using only a threshold comparison does not reduce performance.
Therefore, the qualitative result from the previous section still holds:
with deterministic updates \unit[6]{bit} is enough to achieve the performance level of the baseline simulation.
If updates are performed in a probabilistic manner, \unit[4]{bit} is sufficient to reach the performance of the baseline simulation with added noise.

\paragraph{Effect on weights}
\begin{figure}
  \begin{center}
    \includegraphics[width=\figw]{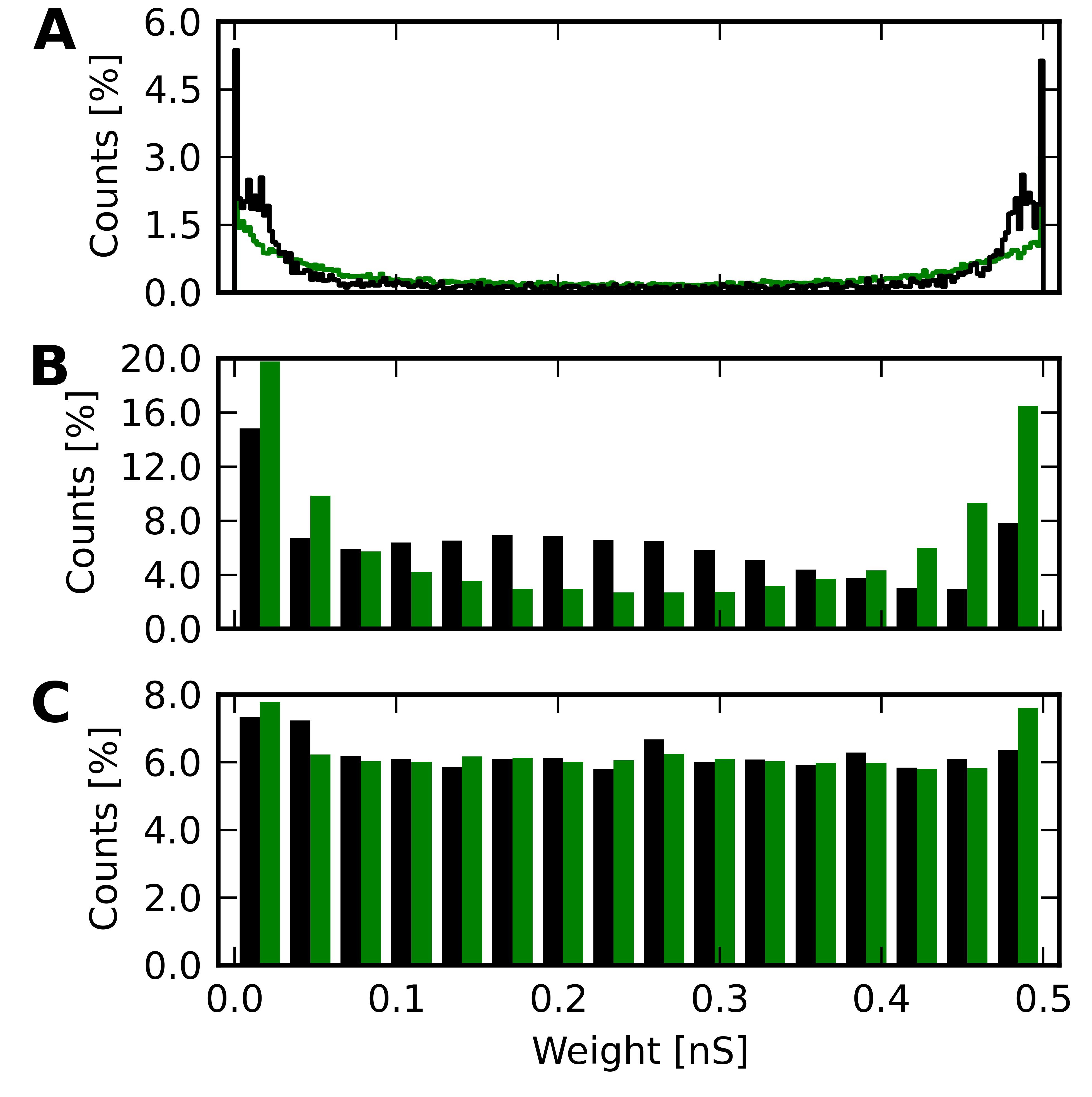}
  \end{center}
  \caption{
    Histogram of synaptic weights after learning with threshold readout.
    \mansubref{A} The histogram is plotted in black for \unit[8]{bit} weights.
    The green histogram shows the result for continuous weights rounded to this resolution.
    \mansubref{B} As in \mansubref{A}, but for \unit[4]{bit} weights with deterministic updating.
    \mansubref{C} Final weights for a resolution of \unit[4]{bit} with probabilistic updating in black.
    Now, the green histogram shows the final weights of the baseline simulation with added noise rounded to this resolution.
  }
  \label{fig:weightdistthresh}
\end{figure}
Comparing the histograms of synaptic weights after learning gives a similar picture to the results of Section~\ref{sec:resdisc}:
With deterministic updates, the histograms have maxima at the upper and lower weight limit as is shown in Figure~\manref{fig:weightdistthresh}{A-B}.
The \unit[4]{bit} case (Figure~\manref{fig:weightdistthresh}{B}) again shows a local maximum around the initial weight value $w_S=\unit[0.21]{nS}$.
In comparison to Figure~\manref{fig:weightdistdisc}{D} this maximum is broader.
With probabilistic updates the histogram is nearly flat (Figure~\manref{fig:weightdistthresh}{C}).
The average root-mean-square error to the mean baseline weights can be compared to the values given in Section~\ref{sec:resdisc}:
For \unit[8]{bit} resolution it is $\left<E_w^\text{8t}\right> = \unit[(0.19\pm 0.06)]{nS}$, which is larger than $\left<E_w^\text{8}\right>$.
For \unit[4]{bit} the error $\left<E_w^\text{4t}\right> = \unit[(0.18\pm 0.11)]{nS}$ is comparable to $\left<E_w^\text{4}\right>$.
With probabilistic updates the result $\left<E_w^\text{4tp}\right> = \unit[(0.15\pm 0.01)]{nS}$ is the same as $\left<E_w^\text{4p}\right>$ without the threshold readout.

The KS test shows larger deviations of the weight distribution for all simulations with deterministic updates compared to having only discrete weights (Table~\ref{tab:performance}).
For \unit[4]{bit} with probabilistic updates the deviation is decreased (Nos.~10 and~6 in Table~\ref{tab:performance}).

\begin{table}
  \centering
  \caption{
    Comparison of simulations with different hardware constraints.
    The table lists the final performance $R_\text{after}$ and the Kolmogorov-Smirnov (KS) measure $D_\text{KS}$ comparing the final weight distribution to the one of the reference simulation indicated by its row number in the last column.
    For this comparison, the continuous reference distribution is rounded to the respective resolution.
    The p-value for the KS test is $p=0.35$ for \unit[8]{bit} and $p<0.01$ in all other cases.
  }
  \label{tab:performance}
  \begin{tabular}{clcccc}
    \toprule
    \textbf{No.} & \textbf{Description} & $R_\text{after}$  & $D_\text{KS}$     & \textbf{Reference} \\
    \midrule
    1  & Baseline                     & $0.54 \pm 0.05$     & -                 & - \\
    2  & Baseline with noise          & $0.45 \pm 0.03$     & -                 & - \\
    \midrule
    \multicolumn{3}{l}{Discretized weights} \\
    \midrule
    3  & \unit[8]{bit}                & $0.53 \pm 0.03$     & $0.008$           & (1) \\
    4  & \unit[6]{bit}                & $0.52 \pm 0.03$     & $0.039$           & (1) \\
    5  & \unit[4]{bit}, deterministic & $0.37 \pm 0.03$     & $0.098$           & (1) \\
    6  & \unit[4]{bit}, probabilistic & $0.46 \pm 0.03$     & $0.053$           & (2) \\
    \midrule
    \multicolumn{3}{l}{Threshold readout} \\
    \midrule
    7  & \unit[8]{bit}                & $0.59 \pm 0.03$     & $0.140$           & (1) \\
    8  & \unit[6]{bit}                & $0.59 \pm 0.05$     & $0.120$           & (1) \\
    9  & \unit[4]{bit}, deterministic & $0.27 \pm 0.04$     & $0.154$           & (1) \\
    10 & \unit[4]{bit}, probabilistic & $0.48 \pm 0.05$     & $0.043$           & (2) \\
    \bottomrule
  \end{tabular}
\end{table}
\subsection{Analog drift}
\label{sec:drift}
\begin{figure}
  \begin{center}
    \includegraphics[width=\figw]{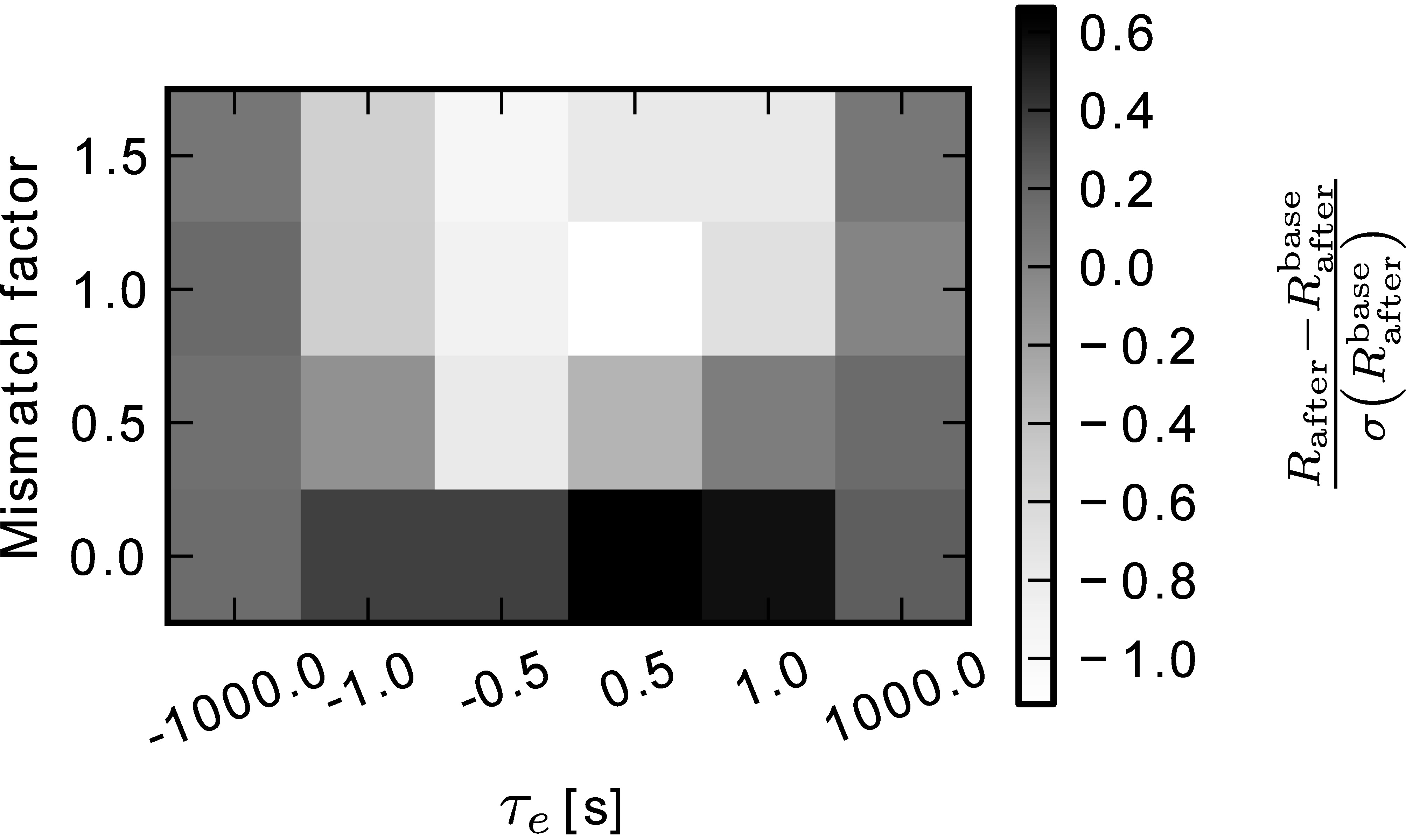}
  \end{center}
  \caption{
    Difference of final reward to the baseline simulation $R_\text{after}-R_\text{after}^\text{base}$ in units of the baseline standard deviation.
    The varied parameters are the average time constant and the amount of mismatch between synapses.
  }
  \label{fig:tauemismatch}
\end{figure}
In the hardware system, the eligibility trace is implemented as an analog variable inside the synapse circuit.
It is therefore subject to drift caused by leakage currents.
In Equation~\ref{eqn:drift}, we have proposed to model this using a drift function.
Additionally, this behavior varies between synapses due to imperfections introduced by the manufacturing process.
This is taken account for by randomly drawing parameters for the drift function from a Gaussian distribution.

To assess the impact of this drift on the performance in the learning task, we performed a sweep over a number of average time constants and degrees of mismatch between synapses.
The results of the simulation, using continuous weights and the thresholded eligibility readout described above, are shown in Figure~\ref{fig:tauemismatch}.
The gray value indicates the difference between $R_\text{after}$ and the baseline value $R_\text{after}^\text{base}$ (Section~\ref{sec:baseline}) in units of the standard deviation of the baseline simulation (darker color is better).
All values fall within one standard deviation of the baseline case, which means that performance is only weakly sensitive to changes of time constant and mismatch of the eligibility trace.
The best performance is achieved for $\tau_e = \unit[0.5]{s}$ and no mismatch ($R_\text{after} = 0.59\pm 0.02$).
In Section~\ref{sec:eligthresh}, the black trace in Figure~\manref{fig:discrete_elig_thresh}{A} shows the reward trace for the same parameters.
The simulation there reached the same performance.
For very large time constants, i.e.\ $\tau_e = \pm \unit[1000]{s}$, drift is negligible compared to the trial duration $t_\text{trial} = \unit[1]{s}$.
This leads to minor deviations in the leftmost ($\left< R_\text{after} \right> = 0.55 \pm 0.02$) and rightmost ($\left< R_\text{after} \right> = 0.55 \pm 0.01$) columns of Figure~\ref{fig:tauemismatch}.
This is above the baseline level, but below the one reached in simulations with threshold readout and \unit[8]{bit} resolution.
The worst performance ($R_\text{after} = 0.45 \pm 0.04$) is obtained for small time constants $\tau_e = \unit[0.5]{s}$ with large mismatch factor $m_e = 1$, because for $\tau_e$ lesser than or equal to the trial duration, the effect of drift is more important.

In this test, the model has shown to be robust to large deviations from the temporal behavior of the eligibility trace in the baseline model.
Drift towards the positive and negative extrema of the eligibility trace, which is the opposite of the desired decaying behavior, does not affect performance.
Neither does variation of up to \unit[150]{\%} of the time constant.
This shows the model to be a well-suited candidate for implementation in neuromorphic hardware, where large variations and distortions are often encountered.

\subsection{Delayed reward}
\begin{figure}
  \centering
  \includegraphics[width=\figw]{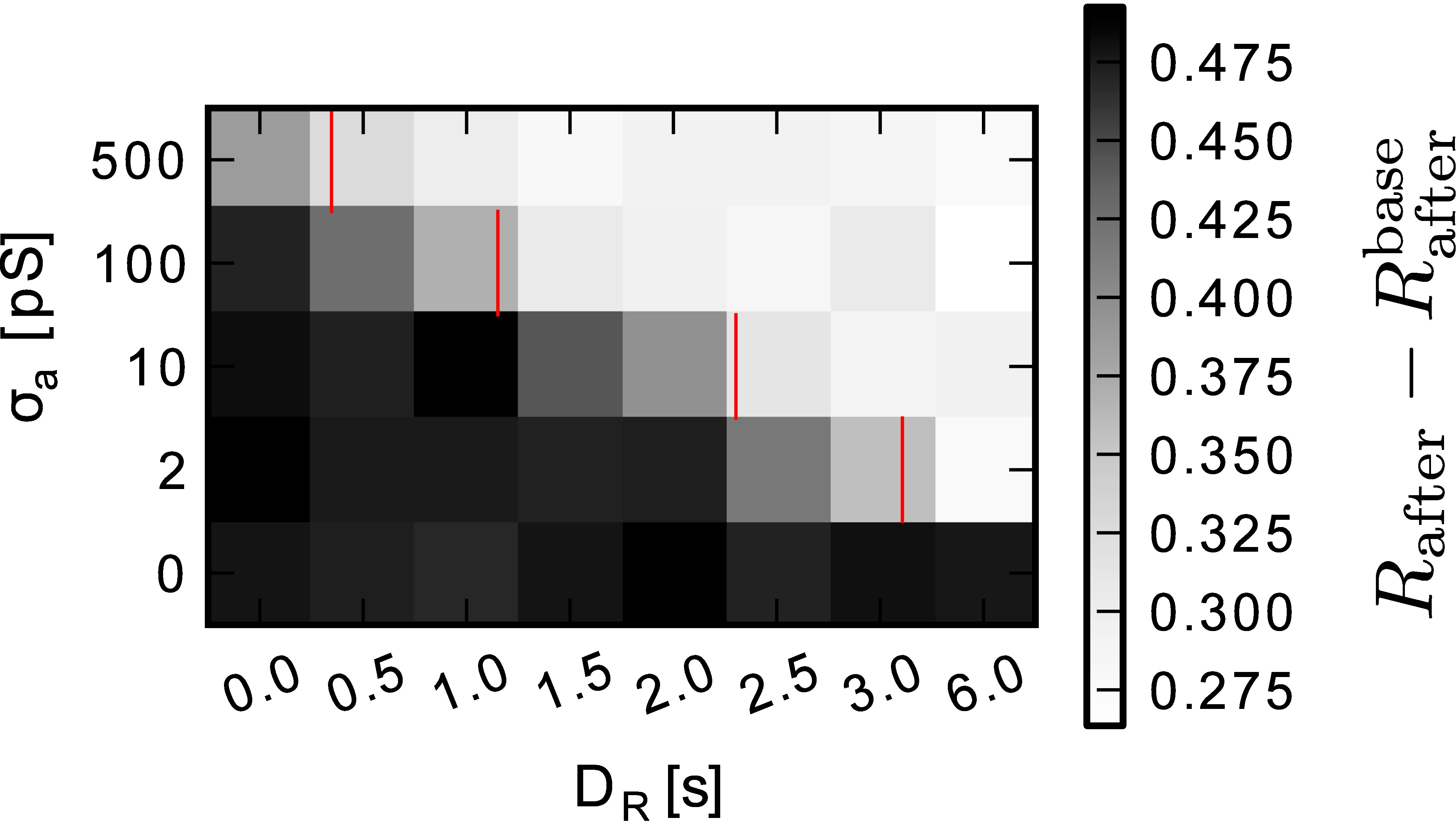}
  \caption{
    Improvement in reward $R_\text{after} - R_\text{before}$ by learning for a range of delays and accumulator readout noise levels.
    Red bars indicate the predicted maximally tolerable delay (Equation~\ref{eqn:dmax3}).
    Data is averaged over 15 simulation runs.
  }
  \label{fig:delayperf}
\end{figure}
In the proposed system, the simulation of the neural network is carried on by analog hardware elements, while the simulation of the environment is left to a conventional computer system.
In this context, latencies due to technical reasons -- e.g., by communication with the environment or computation by the \emp -- can cause temporal delays with respect to ideal calculations.
Additionally, the analog readout of the accumulation traces $a_+, a_-$ is affected by noise.

To better understand the impact of these effects on learning performance, 
a sweep over readout noise and reward latency values was performed, the results of which are shown in Figure~\ref{fig:delayperf}.
The simulation did not include mismatched drift, but used a fixed time constant of \unit[500]{ms} with continuous weights.
The gray value represents the improvement in reward by learning $R_\text{after} - R_\text{before}$.
The data shows that depending on the amount of noise learning is impaired by the delay.
The red bars indicate the predicted maximally tolerable delay assuming a signal-to-noise ratio of one is required (Equation~\ref{eqn:dmax3}).
The simulation fits the prediction well.
A noise level of $\sigma_a = \unit[500]{pS}$ corresponds to \unit[50]{\%} of the maximum of the eligibility trace $a_\text{max}$.

The simulation results confirm that noise on the local accumulation circuit limits tolerable delay.
Because of the accelerated time base of the system, communication delays can easily reach seconds of emulated time.
With an acceleration factor of $\alpha = 10^5$ one second of emulated time is equivalent to $\unit[10]{\mu s}$.
So with \unit[1]{\%} of noise ($\sigma_a = \unit[10]{pS}$), the round-trip-time to the environment must be less than $\unit[20]{\mu s}$ for a $\tau_e = \unit[500]{ms}$ time constant.
Equation~\ref{eqn:dmax3} can be used to find working combinations of the parameters round-trip-time, analog noise and time constant.

\subsection{Towards hardware implementation}
The previous sections have presented results for the performance of the learning rule under various constraints caused by a hardware implementation.
We now want to present simulation results and area estimates for the hardware implementation itself.
So far, the \emp has been produced as an isolated general purpose processor in a \unit[65]{nm} process technology.
A version integrated into the \waferscale was tested in simulation.

The \emp core produced in the \unit[65]{nm} technology covers an area of \unit[0.14]{$\text{mm}^2$} excluding SRAM blocks for \unit[32]{kiB} of main memory.
It was tested using the CoreMark benchmark \citep{coremarkbench} achieving a normalized score of \unit[0.75]{$\frac{\text{Iterations}}{s\cdot \text{MHz}}$}.
At \unit[500]{MHz} and \unit[1.2]{V} supply voltage it consumes \unit[$(48.0 \pm 0.1)$]{mW} of power executing the CoreMark benchmark.

The \waferscale is built in a larger \unit[180]{nm} process technology.
A version with integrated \emp was prepared to estimate area requirements and to simulate the system.
The design was synthesized and standard cell placement was carried out.
This gave an area estimate for the \emp core of \unit[0.895]{$\text{mm}^2$}, excluding the \unit[12]{kiB} of main memory.
All plasticity related logic in the digital part make up \unit[6.2]{\%} of the total design area.
In simulation we tested a weight updating program suitable for the reward modulated STDP rule discussed in this study.
It requires \unit[5.1]{kiB} of main memory and achieves a best-case update rate of \unit[9552]{synapses/s} for \unit[4]{bit} weight resolution.
Due to the lack of hardware support for probabilistic updates and higher weight resolutions than \unit[4]{bit} in the \synapse special-function unit, performance is reduced in these cases.
For probabilistic updates it is \unit[802]{synapses/s} and for \unit[8]{bit} weights \unit[573]{synapses/s}.
Note, that update rates are given in the biological time domain using an acceleration factor of $10^4$.

\section{Discussion}
\label{sec:discussion}
In this study we have proposed a hybrid architecture for plasticity, combining local analog computing with global, program-based processing.
We have then simulated a reward-modulated spike-timing-dependent plasticity learning rule studied by \cite{Fremaux2010} to analyze its implementability.
Starting from a baseline case with no hardware effects, the level of hardware detail of the simulations was increased, with a focus on the negative effects introduced by an implementation using the proposed system.
Note that we did not try to precisely model the hardware device, as it would be done, for example, in a transistor level simulation.
Instead, our goal was to find the effects to which the model is sensitive in order to guide future design decisions.

Overall, we did not find major obstacles for the proposed implementation, but we showed that some design choices are critical to the proper functioning of the learning rule.
In the following, we will discuss guidelines concerning weight resolution, implementation of the eligibility trace and the importance of low-latency communication.
After that, we will consider scalability and flexibility of the approach and compare the design with other hardware systems and discuss the limitations of this study.

\subsection{Weight resolution}
For neuromorphic hardware systems using digitally represented weights, a key question is how many bits to use per synapse, as this determines the amount of wafer area the circuit requires.
For networks with highly connected neurons, small synapses are important for scalability.
This drives implementations to a reduction of the number of bits used for the weight compared to software simulators, which typically use a quasi-continuous \unit[32 or 64]{bit} floating-point representation.
On the other hand, on-line synaptic plasticity learning rules, for example STDP, require incremental changes to the weights.
Discretization confines these changes to a grid with a resolution determined by the number of bits.

For the synaptic plasticity model and the learning task considered, we found that this indeed limits learning performance when using deterministic updates and \unit[4]{bit} weights.
Two solutions to this problem were tested: using higher resolutions and making updates probabilistically.
In the former case, a performance comparable to the continuous case is reached with \unit[6]{bit}.
With probabilistic updates, the performance of \unit[4]{bit} synapses could be improved to nearly the same level.
The comparison to the baseline simulation with added noise of equivalent magnitude showed performance to be limited by the introduced noise and not the discretization of weight values.
Therefore, it is not necessary to build high resolution hardware synapses comparable to software simulators, but even a modest number of bits gives good performance.

In \cite{Seo2011} the authors arrive at a similar result.
They built a completely digital system in a version with \unit[1]{bit} synapses and probabilistic updates and one with \unit[4]{bit} synapses and deterministic updates.
Learning performance in a benchmark task is improved in the latter case, but adds additional costs in area and power consumption.

In \cite{Pfeil2012} the question of weight resolution was also studied for the \waferscale using a synchrony detection task.
Comparable to our findings, they report \unit[8]{bit} weights to perform as good as floating-point weights.
\unit[4]{bit} weights were sufficient for solving the task, but did not reach the same performance.

\subsection{Implementation of the eligibility trace}
In neural models of reinforcement learning, the eligibility trace serves an important purpose:
it allows to connect neural activity with reward.
Reward typically arrives with a delay with respect to the activity underlying causing actions respective spikes.
But only when reward arrives does the agent know how to change the weights.
The hybrid concept of local analog accumulation and global processor-based weight computation fits this model very well.
Therefore, we can identify the local circuit in the synapse with the eligibility trace.
However there are two differences.
First, the processor does not have direct access to the accumulated value, but can only do a simple comparison operation (Equation~\ref{eqn:eval}).
Second, there is no controlled exponential decay of the accumulator.
The analysis in Sections~\ref{sec:eligthresh} and~\ref{sec:drift} shows no degradation in learning performance by both effects.
On the other hand, the lack of controlled and possibly configurable decay presents a constraint to the fidelity, with which learning rules can be implemented.
It is not clear, how other learning tasks would be affected by this lack.


\subsection{Impact of real-world timings}
\label{sec:realworldtimings}
In the presence of delayed reward, three parameters govern whether learning is possible:
1) communication round-trip-time to the environment and back,
2) the amount of noise on the eligibility trace, and
3) the time constant of decay of the eligibility trace.
Equation~\ref{eqn:dmax3} allows to determine working combinations of them.
Reducing the speed-up factor would make communication latency less of a problem, but it would require longer lasting analog storage to achieve the same time constant in emulated time.
Small long-term analog memory is difficult to build due to leakage effects.
Therefore, the triangle of parameters needs to be carefully balanced.
A different approach to deal with communication latency would be to execute the environment on the \emp itself.
This would require adding direct access to spike times by the processor.

\subsection{Scalability and flexibility}
It is important to note, that the synaptic weight and eligibility trace are stored local to the synapse circuit and therefore do not consume processor main memory.
Therefore, for the tested learning rule the required memory does not increase with the number of synapses.
The rule itself can be implemented using \unit[5.1]{kiB} of memory for code and data, which is well below the provided \unit[12]{kiB}.
The time to update all of the synapses scales linearly with their number.
In the proposed hardware system, one \emp processes up to \unit[230]{k} synapses.
Compared to this, the best-case updating rate of \unit[9552]{synapses/s} for the reward modulated STDP rule implies delays on the order of tens of seconds if all synapses are used.
Therefore, the same considerations apply as to the problem of delayed reward discussed in Section~\ref{sec:realworldtimings}.
For the task tested here, simulations indicate no degradation of learning performance for update rates down to \unit[500]{synapses/s} (data not shown).
However, the task only uses a small subset of 1250~synapses.

In general and depending on the task, the updating rate can limit the number of usable plastic synapses per processor to a number below \unit[230]{k}.
This can be met with three strategies:
Randomizing the order of updates, so that over time all synapses are updated with a short delay.
Reducing the acceleration factor by recalibration as long as the resulting neuronal time-constants are still within the achievable range of the circuit.
Distributing plastic synapses over the wafer, so that fewer are used per processor and thereby trading efficiency against fidelity of the emulation.
The last approach is especially suitable if not all synapses in the model require plasticity.

Since the \emp is a general purpose processor, arbitrary C-code can be used to define learning rules.
These rules are restricted by three constraints:
1) The program has to fit into \unit[12]{kiB} of memory.
2) The updating rate establishes a soft limit on the number of plastic synapses per processor.
3) The program can only observe the network activity through the local accumulation circuits.
The last point in particular excludes changing the shape of the STDP curve (Equation~\ref{eqn:stdp}), since it is a fixed property of the local synapse circuit.

Although we only discuss one particular learning rule in detail in this study, a main strength of the system is its ability to implement a wide set of rules.
Going beyond STDP-based rules, two examples would be gradient descent methods and evolutionary algorithms.
In both cases -- as for the STDP rule studied here -- the environment provides a reward signal that guides the change of weights performed locally by the \emp.
For these two examples, the local accumulation circuit is not used at all.
Instead, for gradient descent, or ascent in the case of reward, the gradient of a randomly selected subset of weights is determined by evaluating the performance of the network multiple times and then changing the weights in direction of the gradient.
For evolutionary algorithms, the weights belonging to an individual would be distributed over the wafer, so that every processor has access to a subset of weights of all individuals.
After the reward for each individual is supplied by the environment, the processors can perform combination and mutation on their local subsets in parallel.
Typically, gradient descent and evolutionary algorithms require many evaluations of network performance and are therefore computationally expensive on conventional computers.
In the proposed hardware system, the high acceleration factor, implementation of the network dynamics as physical model, and the parallel weight update promise fast learning with these rules and good scalability with the number of synapses.

\subsection{Comparison to other STDP implementations}
Plasticity implementations found in the literature typically focus on variants of unsupervised STDP and use fixed-function hardware.
For example in \cite{indiveri_tnn2006} STDP works on bi-stable synapses and is implemented using fully analog circuits.
In \cite{ramakrishnan2011floating} analog floating-gate memory is used for weight storage that can be subjected to plasticity.
In contrast, \cite{Seo2011} describes a fully digital implementation using counters and linear-feedback shift registers for probabilistic STDP with single-bit synapses. 
While these systems allow for flexibility, for example in the shape of the timing dependence, there are three main restrictions compared to the processor based implementation presented here:
1) Flexibility is restricted to parameterization of a more or less generic circuit.
2) Weight changes are triggered by spike-events and depend on the timing of spike-pairs.
3) The synapse has no state in addition to the weight.
Points 2) and 3) imply that weights have to be changed immediately in reaction to pre- or postsynaptic spikes.
This rules out the ability to implement an eligibility trace to solve the distal reward problem of reinforcement learning \citep{Izhikevich2007}.

The analog synapse circuit in \cite{Wijekoon2011} does include a local eligibility trace and the ability to modulate the weight update by an external reward signal.
The plasticity of the synapses can be configured to operate under modulation or as unsupervised STDP.
Their approach represents a specialized implementation of reward modulation that emphasizes power and area efficiency.
In contrast, our approach aims for flexibility, so that very different learning rules can be implemented on the same hardware substrate, thereby sacrificing some of the efficiency.
Examples given previously for non-STDP type learning rules are gradient descent and evolutionary algorithms.

However, there are systems that also use a general-purpose processor for plasticity.
For example, in \cite{Vogelstein2003} an implementation of STDP in an address-event representation (AER) routing system is presented.
They use three individual chips:
a custom integrate-and-fire neuron array, an SRAM based look-up table for synaptic connections and a micro-controller for plasticity.
For STDP, the micro-controller processes every spike and maintains queues of pre- and post-synaptic events.
This necessitates multiple off-chip memory accesses for every event and at regular time steps.
Contrary to our approach, their system has access to the detailed timing of spikes and can therefore additionally implement rules including short-term effects, as in \cite{Froemke2010}.
However, in terms of scalability, our proposed system is superior due to the integration of processor, event routing and neuronal dynamics onto the same wafer.
This reduces power consumption by eliminating communication across chip boundaries.
Also, due to the hybrid architecture of analog accumulation and digital weight computation, the workload for the processor is reduced.
This is an important aspect if a high speed-up factor is aimed for.


The system reported in \cite{Davies20123} is a specialized multi-processor platform for neural simulations.
In implementing STDP, a key constraint for them is limited access to weights stored in external memory.
They solve this problem by predicting firing times based on the membrane potential.
This simultaneously illustrates the strength and weakness of this architecture.
Since the system is completely digital, they have unconstrained access to state variables, such as the membrane potential.
With analog neurons, this always requires some form of analog to digital conversion.
On the other hand, weights are stored external to the processor and have to be transfered between chips.
In our system, close integration of weight memory and processor on the same substrate in addition to the optimized input/output instructions of the \synapse special-function unit, make weight access more efficient.


In conclusion, the hybrid processor based architecture proposed in this study represents a novel plasticity implementation for hardware. 
To our knowledge, it introduces two novel concepts: 
first, the integration of a general-purpose processor for plasticity onto the neuromorphic substrate, and
second, the close interaction with specialized analog computational units using an extension of the instruction set.
In combination, this allows for reward-based spike-timing-dependent synaptic plasticity in reinforcement learning tasks.





\subsection{Limitations}
The goal of this study was to analyze the implementability of a reinforcement learning task on a proposed novel hardware system.
The technical implementability of the system itself was not subject of this study.
We assumed a sufficiently fast processor for the delay analysis (Section~\ref{sec:hwconstr}).
It should be part of the design process of a future implementation to test performance against our simulations.
The updating speed could limit the amount of plastic synapses per processor depending on the decay time constant $\tau_e$.
We also did not model the analog part of the system in detail, but restricted simulations to a generic drift function.
Measurements in the existing \waferscale could be used to characterize the drifting behavior.
However, considering that we did not see degraded performance over a large range of time constants and fixed-pattern variation, it does not seem likely that performance would be worse in a more accurate model.

With regard to the model tested here, we restricted the study to one specific task of spike train learning, which is a generic and general learning task for spiking neurons: many tasks can be formulated as a relaxed version of spike train learning.
We showed that the performance of the model is not negatively affected by hardware constraints.
It remains an open question whether there are other tasks that give good performance in software simulations, but fail when hardware constraints are included.
We restricted the study to epochal learning with defined trial-duration ended by the application of the reward.
In a next step, this approach should be extended to continuous time learning scenarios.
In this case, processor update speed and the size of the decay time constant could play a more important role.

\vspace{15pt}


{\bf \large \noindent Conflict of Interest Statement}\\

\noindent The authors declare that the research presented in this paper was conducted in the absence of any commercial or financial relationships that could be construed as a potential conflict of interest.\\
\vspace{15pt}

{\bf \large \noindent Acknowledgments}\\

\noindent The research leading to these results has received funding by the European Union 7th Framework Program  under grant agreement nos.\ 243914 (Brain-i-Nets) and 269921 (BrainScaleS).
We would like to thank Thomas Pfeil for helpful discussions.
\\
\vspace{15pt}

\bibliographystyle{plainnat}
\renewcommand\refname{\subsubsection*{References}}
\small{

}

\end{document}